\documentclass[prd,amsmath,superscriptaddress,twocolumn]{revtex4-1}

\usepackage{hyperref}
\usepackage{ifthen}
\usepackage{amsmath}
\usepackage{amssymb}
\usepackage{graphics}
\usepackage{color,bm}

\newcommand{\tsec}[1]{Sec.~\ref{#1}}
\newcommand{\tsecs}[1]{Secs.~\ref{#1}}
\newcommand{\etal}{\emph{et al}}
\newcommand{\bq}{\begin{equation}}
\newcommand{\eq}{\end{equation}}
\newcommand{\bqa}{\begin{eqnarray}}
\newcommand{\eqa}{\end{eqnarray}}

\newcommand{\rmd}{\ensuremath{\mathrm{d}}}

\newcommand{\Msunh}{M_\odot /h}
\newcommand{\hMpc}{h^{-1}~{\rm Mpc}}
\newcommand{\Mpch}{h~{\rm Mpc}^{-1}}

\newcommand{\Sm}{S_{\rm m}}
\newcommand{\Sbc}{S_{{\rm b}+{\rm c}}}
\newcommand{\Sb}{S_{\rm b}}
\newcommand{\Sc}{S_{\rm c}}
\newcommand{\rhom}{\rho_{\rm m}}
\newcommand{\rhomb}{\bar{\rho}_{\rm m}}
\newcommand{\drhom}{\delta\rhom}
\newcommand{\Om}{\Omega_{\rm m}}
\newcommand{\Olam}{\Omega_{\Lambda}}
\newcommand{\scal}{\varphi}
\newcommand{\varscal}{\phi}
\newcommand{\bscal}{\bar{\scal}}
\newcommand{\dscal}{\delta\varphi}
\newcommand{\dscallin}{\dscal_{\rm lin}}
\newcommand{\absfR}{|f_{R0}|}
\newcommand{\mcom}{\bar{m}}
\newcommand{\mhs}{H_{{\rm m}0}^2}

\newcommand{\Mvir}{M_{\rm vir}}
\newcommand{\rvir}{r_{\rm vir}}
\newcommand{\cvir}{c_{\rm vir}}
\newcommand{\rhos}{\rho_{\rm s}}
\newcommand{\rs}{r_{\rm s}}
\newcommand{\RTH}{r_{\rm th}}
\newcommand{\rhoin}{\rho_{\rm in}}
\newcommand{\rhoout}{\rho_{\rm out}}
\newcommand{\rcham}{r_{\rm c}}
\newcommand{\yhal}{y_{\rm h}}
\newcommand{\yenv}{y_{\rm env}}
\newcommand{\deltac}{\delta_{\rm c}}
\newcommand{\deltacLCDM}{\delta_{\rm c}^{\Lambda}}

\newcommand{\denv}{\delta_{\rm env}}

\begin{document}

\title{Constraining chameleon models with cosmology}

\author{Lucas~Lombriser}

\affiliation{Institute for Astronomy, University of Edinburgh, Royal Observatory, Blackford Hill, Edinburgh, EH9~3HJ, U.K.}

\date{\today}

\begin{abstract}

Chameleon fields may modify gravity on cluster scales while recovering general relativity locally.
This article reviews signatures of chameleon modifications in the nonlinear cosmological structure, comparing different techniques to model them, summarising the current state of observational constraints, and concluding with an outlook on prospective constraints from future observations and applications of the analytic tools developed in the process to more general scalar-tensor theories.
Particular focus is given to the Hu-Sawicki and designer models of $f(R)$ gravity.

\end{abstract}

\maketitle


\section{Introduction} \label{sec:intro}


Attempts to unify general relativity with the standard model interactions typically introduce an effective scalar field in the low-energy limit in addition to the gravitational tensor field.
This fifth element may couple minimally or nonminimally to the matter fields and can act as an alternative to the cosmological constant as explanation for the observed late-time acceleration of our Universe.
A fifth force originating from the nonminimal coupling consequently modifies the gravitational interactions between the matter fields.
Local observations place tight constraints on the existence of a fifth force~\cite{will:05}.
However, these constraints are alleviated for scalar field potentials that yield a scalar field inversely dependent on curvature such as is realised in chameleon models~\cite{khoury:03a, khoury:03b, brax:04, cembranos:05, mota:06}.
In this case, the extra force is suppressed in high-density regions like the Solar System
but at the cost of returning cosmic acceleration to be driven by the contribution of a dark energy component or a cosmological constant rather than originating from a fundamental modified gravity effect of the model~\cite{wang:12}.
Meanwhile, gravitational forces remain enhanced at low curvature and below the Compton wavelength of the scalar field, causing an increase in the growth of structure and rendering nonlinear cosmological structures a vital regime for testing gravity.

This article reviews the signatures of chameleon fields in the cosmological small-scale structure, summarising different techniques to model the formation of these structures and current constraints on the chameleon field amplitude and coupling strength from observations.
Thereby, scalar-tensor models are considered that have a constant Brans-Dicke parameter $\omega$, match the $\Lambda$CDM background expansion history, and exhibit a chameleon suppression mechanism of the enhanced gravitational force in high-density regions.
A special emphasis is given to the Hu-Sawicki~\cite{hu:07a} and designer~\cite{song:06} $f(R)$ models~\cite{buchdahl:70, starobinsky:79, starobinsky:80, capozziello:03, carroll:03, nojiri:03}, to which the scalar-tensor model can be reduced in the case of $\omega=0$.
These models have been particularly well studied with constraints from current and expected from future observations reported, for instance, in Refs.~\cite{chiba:03, zhang:05, amendola:06a, amendola:06b, song:06, amendola:06c, li:07, zhang:07, song:07, hu:07a, jain:07, brax:08, zhao:08, hui:09, smith:09t, schmidt:09a, giannantonio:09, reyes:10, lombriser:10, yamamoto:10, jain:10, motohashi:10, gao:10, wang:10, ferraro:10, thomas:11, davis:11, jain:11, hojjati:11, zhao:11b, gilmarin:11, wojtak:11, lombriser:11b, lam:12a, terukina:12, divalentino:12, jain:12, samushia:12, he:12, hu:12, okada:12, hall:12, simpson:12, albareti:12, lombriser:13a, brax:13a, marchini:13a, vikram:13, abebe:13, lam:13, hellwing:13, upadhye:13, marchini:13b, he:13, hu:13, mirzatuny:13, sakstein:13, zu:13, cai:13, baldi:13, arnold:13, lombriser:13c, brax:13c, terukina:13, hellwing:14, dossett:14, davis:14, munshi:14}.
The currently strongest bounds on the chameleon modifications are inferred from comparing nearby distance measurements in a sample of unscreened dwarf galaxies~\cite{jain:12} as well as from the transition of the scalar field required in the galactic halo to interpolate between the high curvature regime of the Solar System, where the chameleon mechanism suppresses force modifications, and the low curvature of the large-scale structure, where gravity is modified~\cite{hu:07a, lombriser:13c}.
Cosmological observables such as cluster profiles~\cite{lombriser:11b} and abundance~\cite{schmidt:09a, lombriser:10, ferraro:10}, galaxy power spectra~\cite{dossett:14}, redshift-space distortions~\cite{yamamoto:10, okada:12}, or the combination of gas and weak lensing measurements of a cluster~\cite{terukina:13} can be used to place independent and strong bounds on the gravitational modifications.

Local and astrophysical probes yield constraints that are 2-3 orders of magnitude stronger than what is inferred from cosmological observations.
It is worth emphasising, however, that they test the coupling of the scalar field to baryons, whereas cosmological probes typically rely on a coupling of the scalar field to dark matter only, except for the constraints inferred using the cluster gas in Ref.~\cite{terukina:13}.
The separation of coupling strengths between the different matter components can be made explicit in the Einstein frame, in which case cosmological constraints can be regarded as independent of the local and astrophysical bounds.
Furthermore, dark chameleon fields that only couple to dark matter may alleviate problems arising through quantum particle production~\cite{erickcek:13a, erickcek:13b} due to high-energy fluctuations in the early universe or large quantum corrections to the scalar field potential in laboratory environments~\cite{upadhye:12a}.

Hence, cosmological scales remain an interesting regime for constraining potential scalar field contributions.
The chameleon mechanism is a nonlinear effect, however, that complicates the description of the cosmological small-scale structure.
$N$-body simulations of chameleon models provide an essential tool for studying the nonlinear regime of structure formation and the suppression of gravitational modifications~\cite{oyaizu:08a, oyaizu:08b, li:09, li:10, zhao:10b, li:11, puchwein:13, llinares:13b}.
These simulations are, however, computationally considerably more expensive than in the Newtonian scenario.
Hence, for the comparison of theory to observations, allowing a full exploration of the cosmological parameter space involved, the development of more efficient modelling techniques for the cosmological small-scale structure is a necessity.
Different approaches have been proposed based on phenomenological frameworks and fitting functions to simulations~\cite{hu:07b, zhao:10, li:11b, zhao:13}, analytical and numerical approximations~\cite{schmidt:10, pourhasan:11, lombriser:12, terukina:12, terukina:13}, the spherical collapse model~\cite{schmidt:08, borisov:11, li:11a, lombriser:13b, lombriser:13c}, excursion set theory~\cite{li:11a, li:12b, lam:12b, lombriser:13b, kopp:13}, the halo model~\cite{schmidt:08, lombriser:11b, lombriser:13c}, and perturbation theory~\cite{koyama:09, brax:13b}.
These different methods shall be summarised and compared here.


\tsec{sec:chameleonmodels} reviews chameleon gravity in the context of more general scalar-tensor theories and discusses the Hu-Sawicki and designer $f(R)$ models.
\tsec{sec:chameleoncosmology} is devoted to the formation of large-scale structure in chameleon models and its description using linear cosmological perturbation theory in the quasistatic limit, dark matter $N$-body simulations, and the spherical collapse model.
It furthermore discusses different modelling techniques for the matter power spectrum and the properties of chameleon clusters such as the halo mass function and linear halo bias as well as the cluster profiles of the matter density, scalar field, and dynamical mass.
In \tsec{sec:observationalconstraints}, current constraints on chameleon models, in specific on $f(R)$ gravity and from cosmological observations, are summarised.
Finally, \tsec{sec:outlook} provides an outlook and discussion of prospective constraints from future observations and applications of the modelling techniques developed for the chameleon modifications to more general scalar-tensor theories, before \tsec{sec:conclusion} concludes the review.


\section{Chameleon models} \label{sec:chameleonmodels}


The modified gravity models studied here are best viewed as a special limit of the more general scalar-tensor theory defined by the Horndeski extension~\cite{horndeski:74, deffayet:11} to the Einstein-Hilbert action
\bqa
  S & = & \frac{1}{2\kappa^2} \int \rmd^4x \sqrt{-g} \left\{ \vphantom{\frac{0}{0}} G_2(\scal,X) - G_3(\scal,X) \Box\scal \right. \nonumber\\
  & & + G_4(\scal,X) R + \frac{\partial G_4}{\partial X} \left[(\Box\scal)^2-(\nabla_{\mu}\nabla_{\nu}\scal)^2\right] \nonumber\\
  & & + G_5(\scal,X) G_{\mu\nu} \nabla^{\mu}\nabla^{\nu}\scal \nonumber\\
  & & \left. - \frac{1}{6}\frac{\partial G_5}{\partial X} \left[ (\Box\scal)^3 - 3\Box\scal(\nabla_{\mu}\nabla_{\nu}\scal)^2+2(\nabla_{\mu}\nabla_{\nu}\scal)^3\right] \right\} \nonumber\\
  & & + \Sm\left[\psi_{\rm m}; g_{\mu\nu}\right],
  \label{eq:horndeski}
\eqa
where $X\equiv-\frac{1}{2}(\partial_{\mu}\scal)^2$, $R$ is the Ricci scalar, $G_{\mu\nu}$ is the Einstein tensor, $\Sm$ is the matter action with matter fields $\psi_{\rm m}$, $\kappa^2 \equiv 8 \pi \, G$ with the bare gravitational coupling $G$, and natural units are assumed here and throughout the article.
The scalar field $\scal$ is coupled to the metric $g_{\mu\nu}$ via the covariant derivatives, $R$, and $G_{\mu\nu}$.
Eq.~(\ref{eq:horndeski}) defines the most general scalar-tensor theory for which the Euler-Lagrange equations involve at most second order derivatives of the fields.
Hereby, $G_{2-5}$ are free functions of $\scal$ and $X$.

In the following, we specialise to a subclass of the Horndeski action, the Jordan-Brans-Dicke models, which are embedded in Eq.~(\ref{eq:horndeski}) via the definitions
\bqa
 G_2(\varphi,X) & \equiv & -2\left[ U(\varphi)-\frac{\omega(\varphi)}{\varphi}X \right], \nonumber\\
 G_3(\varphi,X) & \equiv & G_5(\varphi,X) \equiv 0, \nonumber\\
 G_4(\varphi,X) & \equiv & \varphi, \label{eq:jordanbransdicke}
\eqa
where $U(\scal)$ is the scalar field potential and $\omega(\scal)$ is the Brans-Dicke parameter determining the kinetic coupling, assumed to be constant in the following with $\omega>-3/2$ to evade ghost fields.
Note that Eq.~(\ref{eq:horndeski}) also embeds, for instance, the Galileon models~\cite{nicolis:08}, which can be represented by the $G_{2-5}$ terms, and the symmetron models~\cite{hinterbichler:10}, which can be represented by appropriate choices of $\omega(\scal)$ and $U(\scal)$.

The Jordan frame action defined by Eqs.~(\ref{eq:horndeski}) and (\ref{eq:jordanbransdicke}) can be recast in the Einstein frame using the transformations
\bqa
 \tilde{g}_{\mu\nu} & \equiv & \scal\,g_{\mu\nu}, \label{eq:metrictransform} \\
 \left(\frac{\rmd \varscal}{\rmd \scal}\right)^2 & \equiv & \frac{1}{2\kappa^2} \frac{3+2\omega}{\scal^2}, \label{eq:scalrel} \\
 A(\varscal) & \equiv & \scal^{-1/2}, \\
 V(\varscal) & \equiv & \frac{U(\scal)}{\kappa^2\scal^2},
\eqa
such that
\bqa
 S & = & \int d^4x \sqrt{-\tilde{g}} \left[ \frac{\tilde{R}}{2\kappa^2} - \frac{1}{2}\partial^{\mu}\varscal\,\partial_{\mu}\varscal - V(\varscal) \right] \nonumber\\
 & & + \Sm\left[\psi_{\rm m}; A^2(\varscal) \tilde{g}_{\mu\nu}\right], \label{eq:einsteinaction}
\eqa
where here and throughout the article, tildes denote quantities in the Einstein frame.

It is worth emphasising that if considering the Einstein frame action Eq.~(\ref{eq:einsteinaction}) to be the fundamental action defining the chameleon model, the fifth element may not be restricted to couple to the different matter components with the same coupling strength, for instance, discriminating between the baryonic components (b) and the cold dark matter (c).
In this case, the matter action in Eq.~(\ref{eq:einsteinaction}) is replaced by $\Sm\left[\psi_{\rm m}; A^2(\varscal) \tilde{g}_{\mu\nu}\right] \rightarrow \Sbc$ with
\bq
  \Sbc = \Sb\left[\psi_{\rm b}; A_{\rm b}^2(\varscal) \tilde{g}_{\mu\nu}\right] + \Sc\left[\psi_{\rm c}; A_{\rm c}^2(\varscal) \tilde{g}_{\mu\nu}\right],
\eq
where one can define the coupling strengths $\beta_{\rm i}$ to the different matter components as $A_{\rm b}(\varscal) = \exp(\beta_{\rm b}\,\kappa\,\varscal)$ and $A_{\rm c}(\varscal) = \exp(\beta_{\rm c}\,\kappa\,\varscal)$.
This underlines the importance of cosmological tests of scalar field couplings as complement to the local tests.
For instance, in the scenario in which $\beta_{\rm b}=0$, the scalar field is minimally coupled to the baryons and Solar System or astrophysical tests as in Refs.~\cite{hu:07a, jain:12, lombriser:13c} do not constrain $\beta_{\rm c}=\beta$, whereas the model can still be constrained using cosmological observations which test the distribution of the cold dark matter.
This review restricts to models in which the coupling strengths to the different matter components are assumed equal, corresponding to $\beta_{\rm b}=\beta_{\rm c}=\beta$ with constant $\beta$ and $\beta\,\kappa\,\phi\ll1$.

The Einstein frame and Jordan frame scalar fields are related by the integration of Eq.~(\ref{eq:scalrel}),
\bq
 \varscal = \frac{1}{\kappa}\sqrt{\frac{3+2\omega}{2}} \ln\scal + \varscal_0,
\eq
where in the following $\varscal_0=0$.
Variation of the action Eq.~(\ref{eq:einsteinaction}) with respect to $\varscal$ yields the scalar field equation
\bq
 \tilde{\Box}\varscal = \frac{\kappa}{\sqrt{6+4\omega}} \tilde{T} + V'(\varscal) \equiv V'_{\rm eff}(\varscal), \label{eq:sfeq}
\eq
where $V_{\rm eff}(\varscal)$ is an effective potential governing the dynamics of $\varscal$ and the energy-momentum tensor is given by $\tilde{T}=A(\varscal)^4 T$.
For a scalar field with $\scal\simeq1$ minimising the effective potential, $V_{\rm eff}'(\varscal)=0$, Eq.~(\ref{eq:sfeq}) becomes
\bq
 \frac{\rmd}{\rmd\scal} U(\scal) \simeq \frac{1}{2}(\kappa^2\rhom+4U)\simeq\frac{\tilde{R}}{2}\simeq\frac{R}{2}, \label{eq:mincond}
\eq
assuming dominance of the matter energy density $\rhom$ and further requiring $(\kappa^2\rhom+4U)\gg(3+2\omega)(\partial_{\mu}\scal)^2/2$.


\subsection{Chameleon mechanism} \label{sec:chameleonmechanism}


With the appropriate choice of $U(\scal)$, the model specified by Eqs.~(\ref{eq:horndeski}) and (\ref{eq:jordanbransdicke}) produces cosmic acceleration and introduces a modification of gravity, which can be suppressed in high-density regions in order to satisfy Solar System constraints.
These demands on $U(\scal)$ can be satisfied with the ansatz $U=\Lambda + U_{\alpha} (1-\scal)^{\alpha}$, where $\alpha$ is a positive constant such that in the limit of $\scal=1$, the model reduces to $\Lambda$CDM.
Hereby, $\Lambda$ may be viewed as an effective cosmological constant, valid as description in the limit of large enough curvature of a more general $U(\scal)$, as is the case in the Hu-Sawicki $f(R)$ models discussed in \tsec{sec:fRgravity}.
On the other hand, chameleon models with equivalent coupling to the different matter fields cannot accommodate cosmic acceleration as a genuine modified gravity effect while simultaneously satisfying Solar System constraints~\cite{wang:12}.
Self-acceleration should be introduced as a consequence of the transformation Eq.~(\ref{eq:metrictransform}) from a non-accelerating expansion in Einstein frame to the Jordan frame, which is not the case here and cosmic acceleration is driven by a cosmological constant as in our ansatz for $U(\scal)$ or a dark energy contribution as, for instance, in the designer $f(R)$ model discussed in \tsec{sec:fRgravity}.

With $\kappa\,\rhom \gg -\sqrt{6+4\omega}\tilde{\nabla}^2\varscal$ in high-density regions, assuming the quasistatic limit, the scalar field is in the minimum of the effective potential $V_{\rm eff}(\varscal)$ in Eq.~(\ref{eq:sfeq}), and hence, $U_{\scal} \simeq R/2$ as in Eq.~(\ref{eq:mincond}), which requires $\alpha\neq1$.
Thus, in this case, and with the effective potential also minimised in the background when $|\omega^j\bscal^{(i)}|\ll1$ with $j=0,1$ and derivatives $i=1,2$~\cite{lombriser:13c}, the corresponding scalar field and its potential become
\bqa
 \scal & \simeq & 1 + (\bscal_0-1) \left(\frac{\bar{R}_0}{R}\right)^{1/(1-\alpha)} \label{eq:scalfield}, \\
 U(\scal) & \simeq & \Lambda -\frac{\bar{R}_0}{2\alpha} \frac{(1-\scal)^{\alpha}}{(1-\bscal_0)^{\alpha-1}}, \label{eq:potential}
\eqa
where subscripts of zero and overbars denote quantities evaluated at present time, $a\equiv1$, and in the background, respectively, here and throughout this review.
Thus, in the high-curvature regime, $R \gg \bar{R}_0$, for $\alpha<1$, the scalar field is suppressed, $(\scal-1) \simeq 2\kappa/\sqrt{6+4\omega}\,\varscal\simeq0$ and consequently, accordingly the gravitational modifications.
In order to have domination of the cosmological constant in the background, $\bar{U}\simeq\Lambda$, and hence, reproduce a $\Lambda$CDM expansion history with $\Delta H^2\sim\mathcal{O}(1-\bscal)$, one further requires $\alpha\gg|\bscal_0-1|$ and $|\omega^j||1-\bscal_0|\ll(1-\alpha)^2$ such that $|\omega^j\scal^{(i)}|\ll1$~\cite{lombriser:13c}.

Local constraints can be estimated by requiring that the Milky Way halo is screened within $8~\textrm{kpc}$, where the Solar System is approximately located, yielding~\cite{lombriser:13c}
\bq
 |\bscal_0-1| \lesssim \frac{5}{6+4\omega} \times 10^{-6} \label{eq:solsyscon}
\eq
with very weak dependence on $\alpha$, which, however, is constrained by the requirement that $\bar{U}_0\approx\Lambda$, implying that $\alpha \gg 10^{-5}(6+4\omega)^{-1}$.

Whereas the chameleon screening mechanism relies on the scalar field mass becoming large in high-density regions, requiring the chameleon model to be valid as an effective field theory with one-loop quantum corrections to $V(\varscal)$ not exceeding the classical scalar field potential, places an upper bound on the mass of the field of the order of $\mathcal{O}(10^{-2})~{\rm eV}$ within a laboratory environment~\cite{upadhye:12a}.
The corresponding minimal scalar field mass set by the Solar System constraint in Eq.~(\ref{eq:solsyscon}), however, is $\gtrsim\mathcal{O}(1-10)~{\rm eV}$.
Hence, the effective field theory interpretation of the models considered in Eq.~(\ref{eq:potential}) is not quantum-stable on these scales.
On the other hand, the classical chameleon models of Eq.~(\ref{eq:potential}) have no effect in E\"ot-Wash type laboratory experiments~\cite{upadhye:12b}.

Besides the Solar System bounds, chameleon models can also be tightly constrained by astrophysical and cosmological probes.
Hereby, it is important to note that in addition to the suppression of modifications in high-density regions due to the chameleon mechanism, on scales larger than the background Compton wavelength today,
\begin{equation}
 \mcom_0^{-1} \sim \sqrt{\frac{3+2\omega}{1-\alpha}\frac{1-\bscal_0}{10^{-6}}}~{\rm Mpc}
\end{equation}
(see \tsec{sec:linearperturbations}), extra forces become Yukawa suppressed.
Thus, with the local constraints in Eq.~(\ref{eq:solsyscon}), this implies that $\mcom_0^{-1}\lesssim(1-\alpha)^{-1/2}~{\rm Mpc}$ and that modified gravity effects are limited to nonlinear cosmological structures (cf.~\cite{wang:12}).

\begin{figure*}
 \centering
 \resizebox{\hsize}{!}{
  \resizebox{0.502\hsize}{!}{\includegraphics{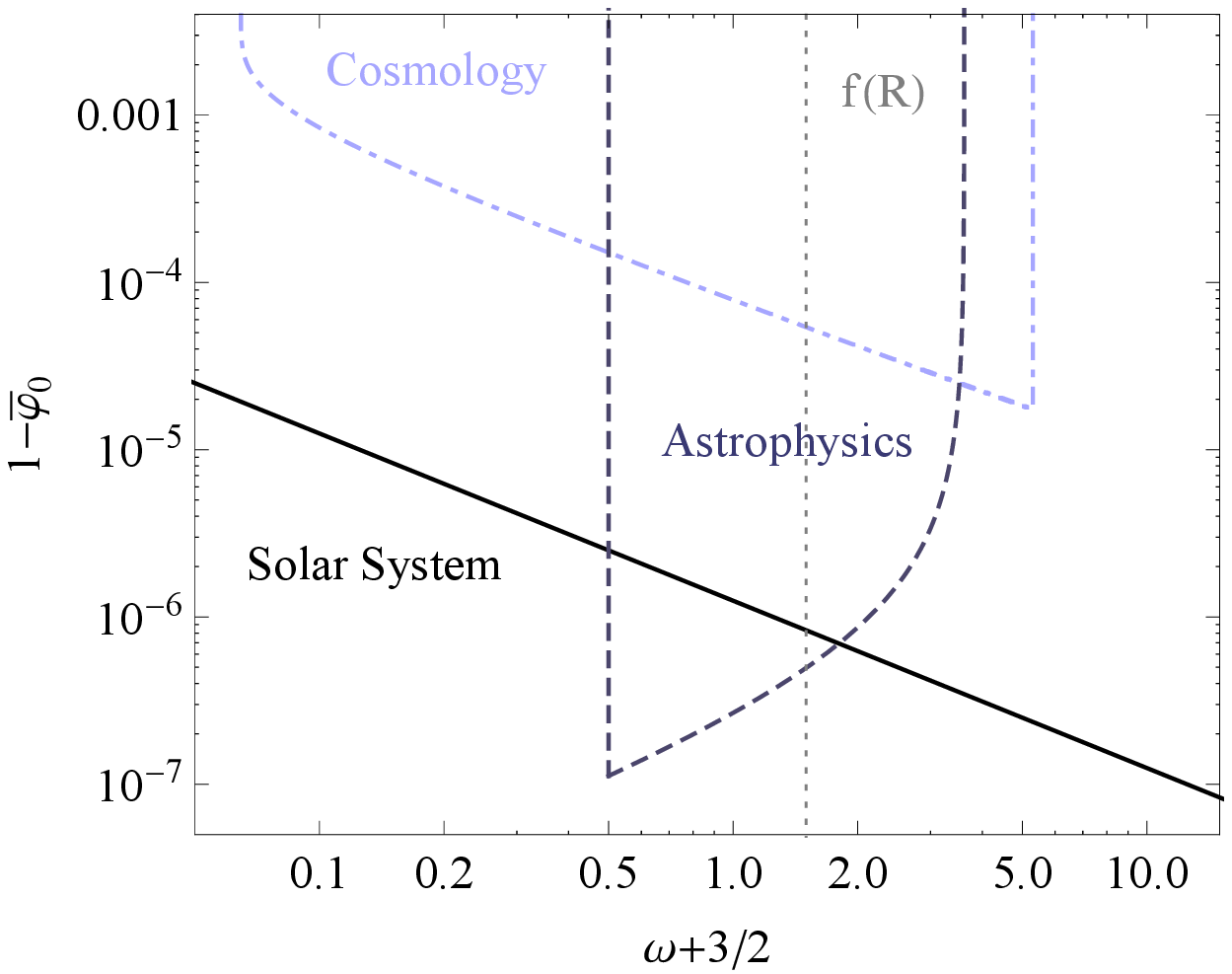}}
  \resizebox{0.498\hsize}{!}{\includegraphics{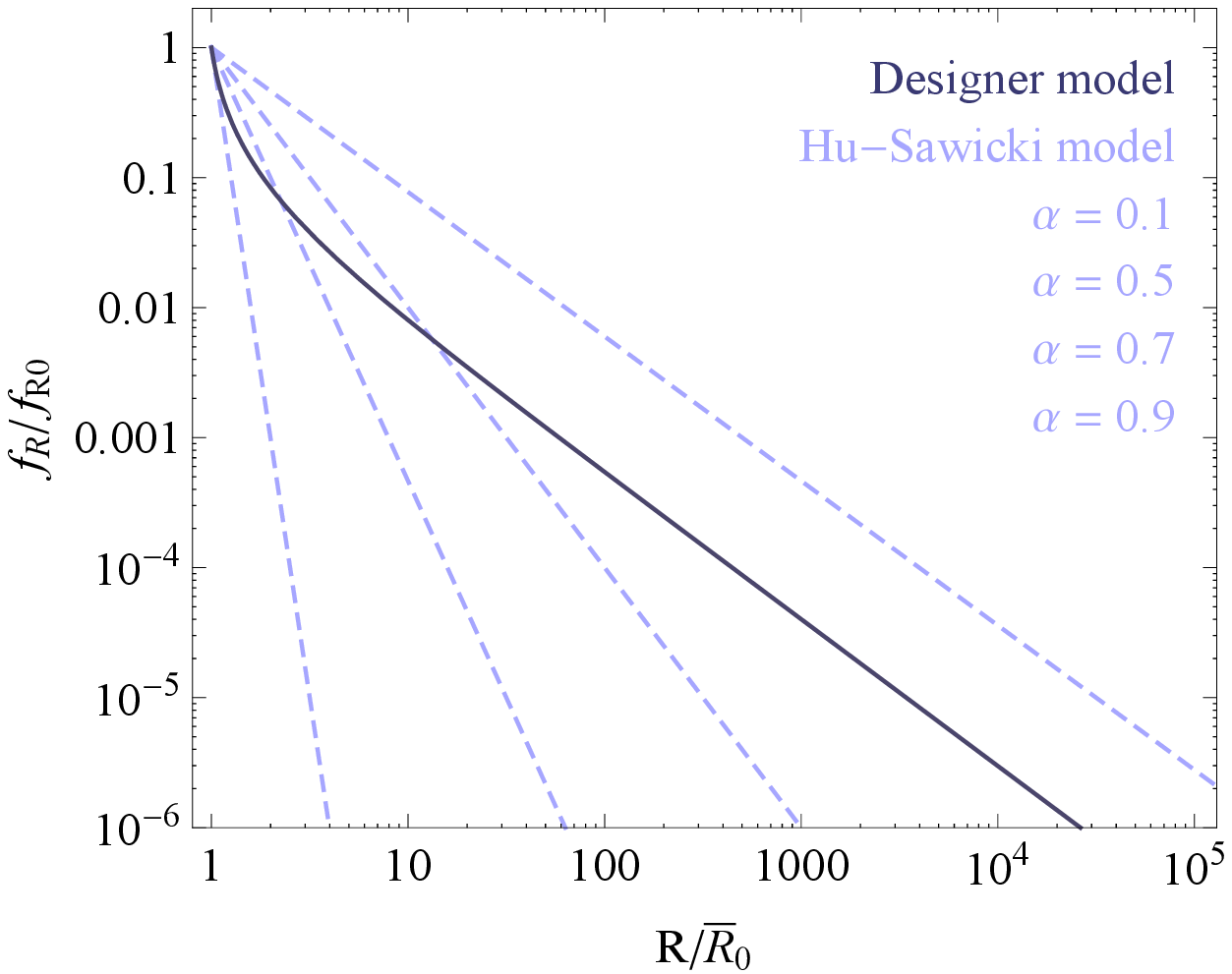}}
 }
 \caption{
  \emph{Left panel:} Local~\cite{lombriser:13c}, astrophysical~\cite{jain:12}, and cosmological~\cite{terukina:13} constraints on the present amplitude of the background chameleon field $\bscal_0$, depending on the Brans-Dicke parameter $\omega$, which is assumed constant and describes the coupling of the chameleon field to matter.
  The vertical dotted line indicates the limit of $f(R)$ gravity.
  \emph{Right panel:} Comparison of the Hu-Sawicki and designer $f(R)$ model scalaron fields.
  }
\label{fig:constraints}
\end{figure*}

Among the strongest constraints on $\bscal_0$ are the bounds that can be inferred from astrophysical observations such as from analysing different distance indicators in unscreened dwarf galaxies~\cite{jain:12} or cosmological tests such as from the comparison of gas to weak lensing measurements in the Coma cluster~\cite{terukina:13}.
Fig.~\ref{fig:constraints} summarises and compares these constraints from the different astronomical regimes, including the local bounds in Eq.~(\ref{eq:solsyscon}), as function of the coupling strength $\omega$ and present background field amplitude $\bscal_0$.
The corresponding constraints in the limit of $f(R)$ gravity (see \tsec{sec:fRgravity}) are indicated by the vertical dotted line.
Note that these results depend on the requirement that $\beta_{\rm b}=\beta_{\rm c}$ and do not apply to scenarios where $\beta_{\rm b}=0$, in which case, however, modifications can, for instance, be constrained through signatures in the observed cluster abundance~\cite{schmidt:09a, lombriser:10, ferraro:10, lombriser:11b}.


\subsection{$f(R)$ gravity} \label{sec:fRgravity}


In $f(R)$ gravity, the Einstein-Hilbert action is supplemented with a free nonlinear function of the Ricci scalar, replacing $R \rightarrow R+f(R)$ in the Lagrangian density~\cite{buchdahl:70, starobinsky:79, starobinsky:80, capozziello:03, carroll:03, nojiri:03}.
This modification can be embedded in the the Jordan-Brans-Dicke subclass of the Horndeski action defined by Eqs.~(\ref{eq:horndeski}) and (\ref{eq:jordanbransdicke}) with $\omega\equiv0$, $\scal\equiv1+\rmd f/\rmd R \equiv 1+f_R$, and scalar field potential $U\equiv(R f_R-f)/2$.
Two particularly well studied classes of $f(R)$ models are the Hu-Sawicki and designer forms. 


The Hu-Sawicki model~\cite{hu:07a} is described by the functional form
\bq
 f(R) = -\mhs \frac{c_1 \left( R/\mhs \right)^n}{c_2 \left( R/\mhs \right)^n + 1}
 \label{eq:husawicki}
\eq
with $\mhs \equiv \kappa^2 \, \bar{\rho}_{{\rm m}0} / 3$.
The parameters $c_1$, $c_2$, and $n$ are the degrees of freedom of the model, which can be set in order to match the $\Lambda$CDM background expansion history and satisfy Solar System constraints through the chameleon suppression mechanism.
For sufficiently large curvature, $c_2^{1/n} R \gg \mhs$, Eq.~(\ref{eq:husawicki}) simplifies to
\bq
 f(R) = -\frac{c_1}{c_2} \mhs - \frac{f_{R0}}{n} \frac{\bar{R}_0^{n+1}}{R^n}
 \label{eq:hsbackground}
\eq
with $f_{R0} \equiv f_R(\bar{R}_0)$.
Requiring equivalence with $\Lambda$CDM when $\absfR \rightarrow 0$ furthermore sets
\bq
 \frac{c_1}{c_2} \mhs = 2\kappa^2 \, \bar{\rho}_{\Lambda}.
\eq
Note that the scalar-tensor model defined by the combination of Eqs.~(\ref{eq:horndeski}) and (\ref{eq:jordanbransdicke}) with the scalar field potential Eq.~(\ref{eq:potential}) reduces to the Hu-Sawicki $f(R)$ gravity model, Eq.~(\ref{eq:hsbackground}), in the limit of $\omega=0$, where $\alpha=n/(n+1)$.
The corresponding normalised scalaron field $f_R/{f_{R0}}$ is shown as a function of $R/\bar{R}_0$ for different values of $\alpha$ in Fig.~\ref{fig:constraints}.
The scalar field mass in the background evaluated today $\mcom_0$ can be related to $\absfR$ following Eq.~(\ref{eq:scalmass}) with $\mcom_0^2\simeq10^{-7}(4-3\Om)(1-\alpha)\absfR^{-1}h^2~\textrm{Mpc}^{-2}$.


Another well studied class of $f(R)$ models are the designer models~\cite{song:06, pogosian:07, nojiri:06a, nojiri:06b, nojiri:10}. 
In this case, $f(R)$ is reconstructed from a predefined background expansion history, which here, shall be given by the matter-dominated $\Lambda$CDM Hubble parameter $H^2=\kappa^2(\rhomb+\bar{\rho}_{\Lambda})/3$.
Using this requirement in the Friedmann equations of $f(R)$ gravity yields an inhomogeneous second-order differential equation for $f(R)$,
\bq
 f'' - \left[ 1 + \frac{H''}{H} + \frac{\bar{R}''}{\bar{R}'} \right] f' + \frac{\bar{R}'}{6 H^2} f = -H_0^2 (1-\Omega_{\rm m}) \frac{\bar{R}'}{H^2}, \label{eq:DDE}
\eq
where if not otherwise specified, primes denote derivatives with respect to $\ln a$ here and throughout the article.
Eq.~(\ref{eq:DDE}) can be solved numerically with the initial conditions
\bqa
 f(\ln a_{\rm i}) & = & A \, H_0^2 a_{\rm i}^p - 6 H_0^2 (1-\Om), \\
 f'(\ln a_{\rm i}) & = & p \, A \, H_0^2 a_{\rm i}^p,
\eqa
where $p = (-7 + \sqrt{73})/4$, set at an initial time $a_{\rm i}\ll1$ in the matter-dominated regime.
Hereby, the amplitude of the decaying mode is set to zero to prevent violations of high-curvature constraints.
The initial growing mode amplitude $A$ then characterises a particular solution in the set of functions $f(R)$ that exactly recover a $\Lambda$CDM background expansion history.
Rather than characterising the solutions by an initial condition $A$ at an arbitrary redshift, they can be defined by $f_{R0}$ as in the Hu-Sawicki model or by the Compton wavelength parameter~\cite{song:06}
\bq
 B = \frac{f_{RR}(\bar{R})}{1+f_R(\bar{R})} \bar{R}' \frac{H}{H'} \label{eq:compton}
\eq
at its present value $B_0 \equiv B(a = 1)$.


Fig.~\ref{fig:constraints} compares the normalised scalaron field $f_R/{f_{R0}}$ of the designer model, using the cosmological parameters defined in \tsec{sec:Nbodysimulations}, with the normalised scalaron field of the Hu-Sawicki model.
From Fig.~\ref{fig:constraints} it is clear that although both models reproduce the $\Lambda$CDM background, one exactly, the other one approximately, the scalar fields do not match beyond $\bar{R}_0$ and the modifications of gravity are similar but not equivalent.
The designer model cannot be described by the Hu-Sawicki model using a constant value of $\alpha$.
Note that in the limit of $B_0=0$ or equivalently $f_{R0}=0$, both the designer and Hu-Sawicki $f(R)$ scenarios reduce to the $\Lambda$CDM model, i.e., including the perturbations to the background.


\section{Chameleon cosmology} \label{sec:chameleoncosmology}


Whereas the chameleon models discussed in \tsec{sec:chameleonmodels} match the $\Lambda$CDM background expansion history and recover general relativity in the Solar System, the cosmological structure on scales of a few megaparsecs and less remains modified.
Importantly, this restriction of gravitational modifications to small-scale cosmology is inferred from the local bounds on the coupling of the scalar field to baryons.
Cosmological probes, however, typically test the distribution of dark matter.
Hence, besides providing independent constraints on the same gravitational model at different scales of our Universe, they may also serve as incompensable tests of the presence of scalar fields that couple nonminimally to the dark matter but may not couple equally strong to the baryonic components.
Thus, it is important to study the signatures of chameleon fields on the large-scale structure even if the amplitudes and coupling strengths for the particular models considered are, in principle, ruled out by local constraints. 

\tsec{sec:linearperturbations} begins with reviewing the linear growth of structure in Jordan-Brans-Dicke gravity and the resulting linear matter power spectrum.
The description of cosmological structure in the nonlinear regime is then discussed in \tsec{sec:Nbodysimulations} using $N$-body simulations and in \tsec{sec:sphericalcollapsemodel} using the spherical collapse model. 
\tsec{sec:chameleonclusters} reviews different approaches of modelling the chameleon effect on halo properties such as in the halo mass function and linear halo bias as well as in the density, scalar field, and mass profiles of the clusters.
Finally, \tsec{sec:matterpowerspectrum} focuses on the description of the nonlinear matter power spectrum, comparing different modelling approaches, including nonlinear parametrisations and fitting functions, the halo model decomposition, and perturbation theory.


\subsection{Linear perturbations} \label{sec:linearperturbations}


In scalar-tensor theories, the linear growth function for matter fluctuations $D_{\scal}(a,k)$ becomes scale dependent and in the quasistatic limit can be determined from solving
\bq
 D_{\scal}'' + \left[ 2 - \frac{3}{2}\Om(a) \right] D_{\scal}' - \frac{3}{2} \mu(a,k) \Om(a)D_{\scal} \simeq 0, \label{eq:modlingroweq}
\eq
where $\Om(a) \equiv H_0^2\Om a^{-3}/H^2$.
Hereby, $\mu(a,k)$ describes the gravitational modification introduced in the Poisson equation by the scalar field, which for the Jordan-Brans-Dicke models defined by Eqs.~(\ref{eq:horndeski}) and (\ref{eq:jordanbransdicke}) takes the form~\cite{esposito:00, tsujikawa:08a}
\bq
 \mu(a,k) = \frac{1}{\bscal}\left[1+\frac{1}{3+2\omega}\frac{k^2\bscal}{a^2 \mcom^2 + k^2\bscal}\right], \label{eq:muak}
\eq
where $\bscal\simeq1$ and for Eq.~(\ref{eq:potential}),
\bq
 \mcom^2 \simeq \frac{1-\alpha}{3+2\omega}\frac{(1-\bar{\scal})^{\alpha-2}}{(1-\bar{\scal}_0)^{\alpha-1}}\bar{R}_0,
 \label{eq:scalmass}
\eq
or $\mcom^2\simeq [3 f_{RR}(R=\bar{R})]^{-1}$ in $f(R)$ gravity.
Eq.~(\ref{eq:modlingroweq}) can be solved setting the initial conditions $D_{\scal}(a_{\rm i},k)=D(a_{\rm i})=a_{\rm i}$ and $D_{\scal}'(\ln a_{\rm i},k)=D'(\ln a_{\rm i})=a_{\rm i}$ at an initial scale factor $a_{\rm i} \ll 1$ in the matter-dominated era.

Assuming the same initial conditions for the chameleon and $\Lambda$CDM models, the modified linear matter power spectrum $P_{{\rm L}\scal}$ relates to the $\Lambda$CDM power $P_{{\rm L}\Lambda{\rm CDM}}$ as
\bq
 P_{{\rm L}\scal}(a,k) = \left( \frac{D_{\scal}(a,k)}{D(a)} \right)^2 P_{{\rm L}\Lambda{\rm CDM}}(a,k).
\eq
The $\Lambda$CDM variance is
\bq
 S(a,r) \equiv \sigma^2(a,r) = \frac{D^2(a)}{D^2(a_{\rm i})} \int  \rmd^3\mathbf{k} \, |\tilde{W}(k\,r)|^2 P_{\rm i}(a_{\rm i},k), \label{eq:variance}
\eq
where $\tilde{W}(k\,r)$ is the Fourier transform of a top-hat function of radius $r$ and $P_{\rm i}$ denotes the power spectrum at initial time $a_{\rm i}$.
Rather than defining the variance at radius $r$, it can also be written as a function of the mass $M=4\pi\,\rhomb\,r^3/3$ enclosed by the top hat.

Beyond the (reduced) Compton wavelength of the background field, $\lambda_{\rm C}\equiv \mcom^{-1}$ from Eq.~(\ref{eq:scalmass}), Eq.~(\ref{eq:muak}) becomes $\mu(a,k)\rightarrow1$ and Eq.~(\ref{eq:modlingroweq}) reduces to the ordinary differential equation for the scale-independent growth function of matter fluctuations $D(a)$ in $\Lambda$CDM.
Note that whereas for the chameleon models, Eqs.~(\ref{eq:modlingroweq}) and (\ref{eq:muak}) strictly only apply in the quasistatic limit, in $\Lambda$CDM with $\mu(a,k)=1$, Eq.~(\ref{eq:modlingroweq}) becomes exact on all scales and can be derived from combining the linearly perturbed Einstein field equations with the energy-momentum conservation in the total matter gauge.
For the scalar-tensor theories with $\Lambda$CDM expansion history considered here, however, deviations in $D_{\scal}(a,k)$ from solving the growth function using the full linear perturbation theory are small~\cite{lombriser:10, hojjati:12, lombriser:13a, lima:13, noller:13}.
Moreover, the following discussion concentrates on the high-curvature regime where Eqs.~(\ref{eq:modlingroweq}) and (\ref{eq:muak}) are accurate but it is worth keeping in mind that in general, in modified gravity models, corrections to the quasistatic approximation may have measurable signatures on horizon scales~\cite{lombriser:11a, lombriser:13a}.


\subsection{$N$-body simulations of chameleon $f(R)$ gravity} \label{sec:Nbodysimulations}


$N$-body simulations are essential in the study of the chameleon mechanism and its effects on the nonlinear cosmological structure.
For this purpose, $N$-body codes have been developed by Refs.~\cite{oyaizu:08a, oyaizu:08b, li:09, li:10, zhao:10b, li:11, puchwein:13, llinares:13b}.
Thereby, the chameleon suppression has been particularly well studied in the case of the Hu-Sawicki $f(R)$ model, for which in the quasistatic limit, the scalar field and Poisson equations become
\begin{eqnarray}
\nabla^2 \delta f_R & = & \frac{a^2}{3} \left[ \delta R (f_R) - 8 \pi \, G \, \delta \rho_{\rm m} \right], \label{eq:fRsim} \\
\nabla^2 \Psi & = & \frac{16 \pi \, G}{3} a^2 \delta\rho_{\rm m} - \frac{a^2}{6} \delta R (f_R), \label{eq:potsim} 
\end{eqnarray}
respectively, where coordinates are comoving, $\Psi = \delta g_{00} / (2g_{00})$ is the Newtonian potential in the longitudinal gauge, $\delta f_R = f_R(R) - f_R(\bar{R})$, $\delta R = R - \bar{R}$, and $\delta \rho_{\rm m} = \rho_{\rm m} - \bar{\rho}_{\rm m}$.
The first simulations solving Eqs.~(\ref{eq:fRsim}) and (\ref{eq:potsim}) were performed by Oyaziu \etal.~\cite{oyaizu:08a, oyaizu:08b} with a particle-mesh code on regular grids.
The resolution of the chameleon force enhancement was then improved using a self-adaptive grid structure by Zhao \etal.~\cite{zhao:10b} with {\sc mlapm}~\cite{knebe:01} and by Li \etal.~\cite{li:12c} with {\sc ecosmog}~\cite{li:11} that is based on {\sc ramses}~\cite{teyssier:01}.
Puchwein \etal.~\cite{puchwein:13} introduced {\sc mggadget}, a modification of {\sc gadget}~\cite{springel:05}, using its tree structure to improve the resolution.
The equations of motion, Eqs.~(\ref{eq:fRsim}) and (\ref{eq:potsim}), in all of these simulations are solved in the quasistatic limit, which is a good approximation for $f(R)$ gravity~\cite{oyaizu:08a, llinares:13b, noller:13}.
Llinares and Mota~\cite{llinares:13a} introduced a non-static solver for the equations of motion in the symmetron model, which they apply in its static version in the {\sc ramses} based $N$-body code {\sc isis}~\cite{llinares:13b} to simulate the Hu-Sawicki $f(R)$ model.

The cosmological structure formed in these simulations has been analysed through measurements of the matter power spectrum~\cite{oyaizu:08b, schmidt:08, zhao:10b, li:11, li:12c, llinares:13b, puchwein:13}, halo mass function~\cite{schmidt:08, ferraro:10, zhao:10b}, halo bias~\cite{schmidt:08}, halo density profile~\cite{schmidt:08, lombriser:11b, lombriser:12}, halo concentration~\cite{lombriser:12}, velocity dispersion~\cite{schmidt:10, lam:12a, lombriser:12, arnold:13}, velocity divergence power spectrum~\cite{jennings:12, li:12c}, matter velocity dispersion cross power~\cite{jennings:12, li:12c}, redshift-space distortions~\cite{li:12c}, gravitational redshift profiles~\cite{gronke:13}, abundance of subhalos~\cite{arnold:13}, and the integrated Sachs-Wolfe effect~\cite{cai:13}.
Moreover, hydrodynamical simulations of $f(R)$ gravity have been performed with {\sc mggadget} to analyse degeneracies appearing between the signatures of $f(R)$ modifications of gravity and baryonic processes on the matter power spectrum, also examining the impact from active galactic nuclei feedback~\cite{puchwein:13}.
Further degeneracies with signatures of massive neutrinos on the simulated matter power spectrum, halo mass function, and halo bias have been discussed in Ref.~\cite{baldi:13}.
Other studies analysed the hydrodynamics of the intracluster and intragroup medium~\cite{arnold:13} or combined simulations with a semi-analytic model for the study of galaxy evolution~\cite{fontanot:13}.

The impact of $f(R)$ modifications of gravity on the halo mass function and matter power spectrum is discussed in \tsecs{sec:chameleonclusters} and \ref{sec:matterpowerspectrum}.
For the numerical computations, also involving the spherical collapse density in \tsec{sec:sphericalcollapsemodel}, the cosmological parameters are set to $\Om=1-\Olam$ with $\Olam=0.76$, $\Omega_{\rm b}=0.04181$, dimensionless Hubble constant $h=0.73$, slope of the primordial power spectrum $n_{\rm s}=0.958$, and initial power in curvature fluctuations $A_{\rm s}$ such that the power spectrum normalisation is $\sigma_8\equiv\sigma(a=1,r=8\hMpc)=0.8$ for $\Lambda$CDM.
This is in correspondence to the settings in the $N$-body simulations of Ref.~\cite{li:12c, zhao:10b}, from which results for the halo mass function and matter power spectrum are also shown in \tsecs{sec:chameleonclusters} and \ref{sec:matterpowerspectrum}, respectively.
The same simulation output has been used as comparison for the analytical modelling of these observables in Ref.~\cite{lombriser:12, lombriser:13b, lombriser:13c}.
Here, in order to highlight the effects of modified gravity and the chameleon mechanism, numerical computations are specialised to $|\bscal_0-1|=\absfR=10^{-5}$ with exponent $\alpha=1/2$ ($n=1$), for which the scalar field is small enough to demonstrate the nonlinear chameleon suppression effect yet large enough to yield a significant modification of the cosmological structure.
Note that analogous to Ref.~\cite{zhao:10b}, halos will be defined using the $\Lambda$CDM virial overdensity $\Delta_{\rm vir}\approx390$, which, motivated by an equal-overdensity approach for the comparison of the structures produced, is also applied to identify $f(R)$ halos.
Hence, the associated virial masses $\Mvir=4\pi\rhomb\Delta_{\rm vir}r_{\rm vir}^3/3$ do not strictly represent virial masses in $f(R)$ gravity~\cite{schmidt:08, lombriser:13b}.

Finally, it is important to note that Eqs.~(\ref{eq:fRsim}) and (\ref{eq:potsim}) form a highly nonlinear system of differential equations, which is harder and takes longer to solve than performing $N$-body simulations in Newtonian gravity.
Additionally, the Hu-Sawicki $f(R)$ model is only a subset of scalar-tensor theories or modifications of gravity of cosmological interest.
Hence, it is important to develop simpler, (semi)analytic tools that allow a more general analysis of the effects on the nonlinear structure caused by modifying gravity.
These are ideally also efficient enough for the application in Markov Chain Monte Carlo explorations of the associated parameter spaces, comparing these computations to observations.


\subsection{Spherical collapse model} \label{sec:sphericalcollapsemodel}



\begin{figure*}
 \centering
 \resizebox{\hsize}{!}{
  \resizebox{0.5\hsize}{!}{\includegraphics{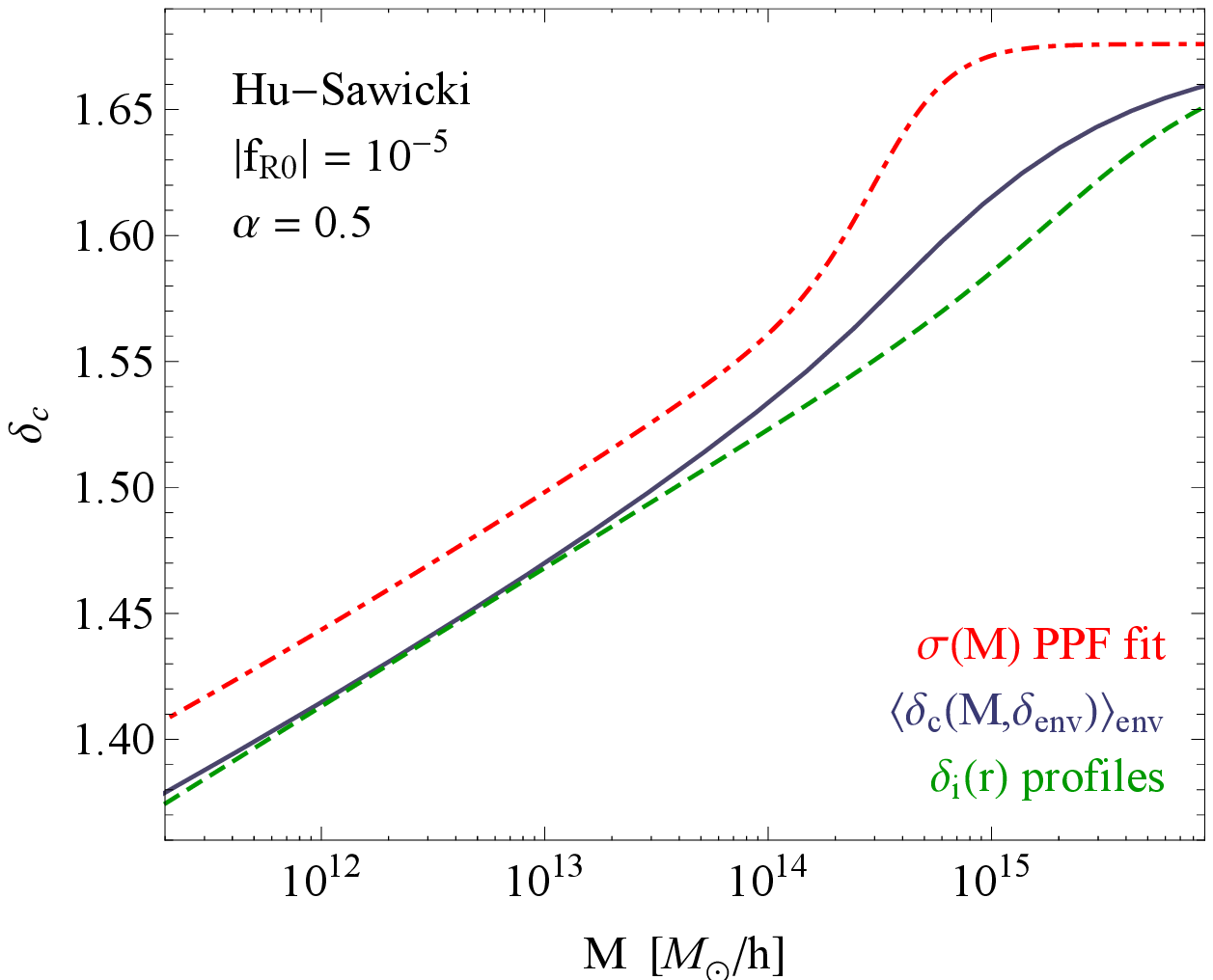}}
  \resizebox{0.5\hsize}{!}{\includegraphics{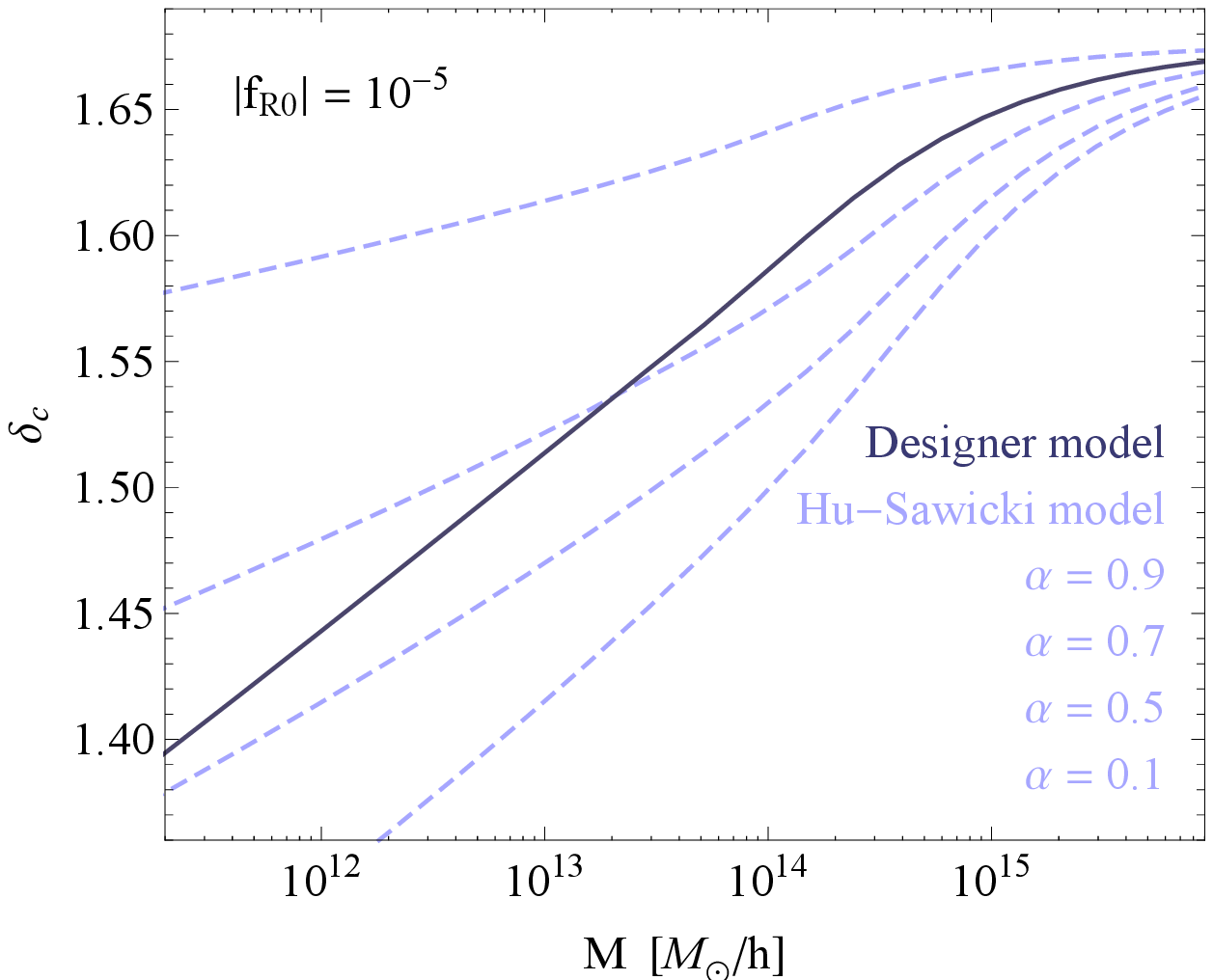}}
 }
 \caption{
  \emph{Left panel:} Comparison between different approaches to compute the spherical collapse density at $z=0$ in the Hu-Sawicki $f(R)$ model with $\alpha=0.5$ $(n=1)$ and $\absfR=10^{-5}$: environmental average of collapse densities from the thin-shell force implementation in the spherical collapse in Eqs.~(\ref{eq:enhf}) and (\ref{eq:thshsphcoll}) (solid curve); evolution of initial overdensity profile (dashed curve); reconstructed collapse density from a phenomenological chameleon transition in the variance (dot-dashed curve).
  Note that in the mass and environment dependent spherical collapse, the coefficient in Eq.~(\ref{eq:thshsphcoll}) depends on $\absfR/M^{2/3}$, hence, $\deltac$ can be scaled to other values of $\absfR$ by a redefinition of mass.
  \emph{Right panel:} Comparison between the spherical collapse densities $\langle\deltac\rangle_{\rm env}$ at $z=0$ in the designer (solid curve) and Hu-Sawicki (dashed curves) $f(R)$ model for different values $\alpha$.
  }
\label{fig:sphcoll}
\end{figure*}


In addition to running $N$-body simulations of chameleon gravity, the nonlinear structure formation can also be studied in the spherical collapse model.
Hereby, a dark matter halo is approximated by a constant spherically symmetric top-hat density $\rhoin$ of radius $\RTH$, that is embedded in an outer matter density $\rhoout$.
The densities are then evolved according to the nonlinear continuity and Euler equations from an initial time to the time of collapse of $\rhoin$.

In the inner part of the top hat and outside of it, the scalar field is in the minimum of the effective Einstein frame scalar field potential $V_{\rm eff}(\varscal)$ in Eq.~(\ref{eq:sfeq}) with the Jordan frame scalar field values $\scal_{\rm in}$ and $\scal_{\rm out}$ denoting the corresponding solutions for $U_{\scal} \simeq R/2$, for instance, given by Eq.~(\ref{eq:scalfield}).
Khoury and Weltman~\cite{khoury:03b} estimate the distance $\Delta r\geq0$ that the chameleon field $\scal\simeq1$  requires to settle from $\scal_{\rm out}$ to $\scal_{\rm in}$ as
\bq
 \frac{\Delta r}{\RTH}
 \simeq (3+2\omega) \frac{\scal_{\rm in} - \scal_{\rm out}}{\kappa^2 \rho_{\rm in}\RTH^2}. \label{eq:thinshell}
\eq
The force enhancement $\Delta F \equiv F-F_{\rm N}$ at $\RTH$ due to the nonminimally coupled scalar field, relative to the corresponding Newtonian force $F_{\rm N} = G\,m_{\rm t}M/\RTH^2$ with test mass $m_{\rm t}$, can then be approximated by~\cite{li:11a, lombriser:13b, lombriser:13c}
\bq
 \frac{\Delta F}{F_{\rm N}} \simeq \frac{1}{3+2\omega}\min\left[ 3\frac{\Delta r}{\RTH} - 3\left(\frac{\Delta r}{\RTH}\right)^2 + \left(\frac{\Delta r}{\RTH}\right)^3, 1 \right]. \label{eq:enhf}
\eq
This result follows from computing the intermediate scalar field within $r\in[r_0,\RTH]$ in the thin-shell regime $\Delta r=\RTH-r_0\ll \RTH$, which reproduces the force enhancement in the thick-shell regime $\Delta r>\RTH$ when $r_0\ll\RTH$.
Hence, Eq.~(\ref{eq:enhf}) gives an interpolation between the regime of chameleon suppression $\Delta F=0$ and the regime where the total force $F$ is maximally modified, $\Delta F/F_{\rm N}=(3+2\omega)^{-1}$, which is $\mathcal{C}^0$ for $\Delta r/\RTH\rightarrow0$ and $\mathcal{C}^2$ for $\Delta r/\RTH\rightarrow1$.
Note that the force modification in Eq.~(\ref{eq:enhf}) is both dependent on the mass of the halo, $M \equiv 4\pi \, \rhoin \RTH^3/3$, and its environmental density $\rhoout$.

Li and Efstathiou~\cite{li:11a} introduce this force enhancement in the evolution of the spherical shell to study the spherical collapse in chameleon models (cf.~\cite{schmidt:08, brax:10, borisov:11}).
The equation of motion for the physical radius of the spherical top-hat overdensity $\zeta(a)$ then becomes~\cite{schmidt:08, li:11a, lombriser:13b}
\bq
 \frac{\ddot{\zeta}}{\zeta} \simeq -\frac{\kappa^2}{6} \left( \rhomb - 2\bar{\rho}_{\Lambda} \right) - \frac{\kappa^2}{6} \left(1 + \frac{\Delta F}{F_{\rm N}} \right) \drhom
 \label{eq:shellmotion}
\eq
with dots denoting cosmic time derivatives.
At initial time, $a_{\rm i}\ll1$, the physical radius of the spherical shell is given by $\zeta(a_{\rm i})=a_{\rm i}\RTH$ but at later times, $a>a_{\rm i}$, the nonlinear evolution causes $\zeta(a)$ to deviate from this simple linear relation.
This deviation can be characterised by the dimensionless variable $y\equiv \zeta(a)/(a\,\RTH)$, where $\rhom/\rhomb = y^{-3}$ follows from conservation of the mass enclosed in the overdensity,  $\rhomb a^3 \RTH^3 = \rhom \zeta^3$.
Hence, Eq.~(\ref{eq:shellmotion}) can be rewritten as
\bq
 \yhal'' + \left[ 2 - \frac{3}{2} \Om(a) \right] \yhal' + \frac{1}{2} \Om(a) \left(1 + \frac{\Delta F}{F_{\rm N}} \right) \left( \yhal^{-3} - 1 \right) \yhal = 0 \label{eq:yhal}
\eq
with $\rhoin/\rhomb=\yhal^{-3}$ and
\bq
 \yenv'' + \left[ 2 - \frac{3}{2} \Om(a) \right] \yenv' + \frac{1}{2} \Om(a) \left( \yenv^{-3}-1 \right) \yenv = 0, \label{eq:yenv}
\eq
for the environment with $\rhoout/\rhomb=\yenv^{-3}$, which follows a $\Lambda$CDM evolution, corresponding to Eq.~(\ref{eq:shellmotion}) in the limit of $\Delta F\rightarrow0$.
The force enhancement in Eq.~(\ref{eq:yhal}) is obtained from Eq.~(\ref{eq:enhf}), replacing $\Delta r / \RTH \rightarrow \Delta\zeta/\zeta$, where for the chameleon models defined by Eq.~(\ref{eq:potential}),
\bqa
 \frac{\Delta \zeta}{\zeta} & \simeq & \frac{(3+2\omega)(\bscal_0-1) \, a^{\frac{4-\alpha}{1-\alpha}}}{3\Om(H_0\RTH)^2} \yhal \left[ \left( \frac{1+4\frac{\Olam}{\Om}}{\yhal^{-3} + 4\frac{\Olam}{\Om} a^3} \right)^{\frac{1}{1-\alpha}} \right. \nonumber\\
 & & \left. - \left( \frac{1+4\frac{\Olam}{\Om}}{\yenv^{-3} + 4\frac{\Olam}{\Om} a^3} \right)^{\frac{1}{1-\alpha}} \right]. \label{eq:thshsphcoll}
\eqa
This system of differential equations is then solved with the initial conditions
\bq
 y_{\rm h/env, i} = 1 - \frac{\delta_{\rm h/env, i}}{3}, \ \ \ \ \ y_{\rm h/env, i}' = - \frac{\delta_{\rm h/env, i}}{3},
\eq
set at an initial scale factor $a_{\rm i} \ll 1$ in the matter-dominated regime.

In order to evade problems with the scale-dependent linear growth in scalar-tensor theories $D_{\scal}(a,k)$ (see \tsec{sec:linearperturbations}), the initial overdensities $\delta_{\rm h, i}$ and $\delta_{\rm env, i}$ can be extrapolated to $a>a_{\rm i}$ using the $\Lambda$CDM linear growth function $D(a)$ in \tsec{sec:linearperturbations} to define an effective linear overdensity
\bq
 \delta_{\rm h/env}({\bf x}; \zeta_{\rm h/env}) \equiv \frac{D(a)}{D(a_{\rm i})} \delta_{\rm h/env, i}. \label{eq:extrapolation}
\eq
In specific, using Eq.~(\ref{eq:extrapolation}) to extrapolate the initial overdensities yielding collapse at a given redshift, defines the effective linear collapse density $\deltac$ and the environmental density $\denv$.
Analogous to these two definitions, in Eq.~(\ref{eq:variance}), the variance $S$ has been defined by the integration of the initial matter power spectrum extrapolated to $a$ using the $\Lambda$CDM linear growth function $D(a)$.
Thus, the peak threshold, which will be of interest in \tsec{sec:halomassfunctionlinearhalobias}, is determined by $\nu=\deltac/\sqrt{S}=\delta_{c,\rm i}/\sqrt{S_{\rm i}}=\nu_i$ due to the scale-independent growth of structure in $\Lambda$CDM.
In contrast, if using the scale-dependent linear growth function $D_{\scal}(a,k)$ of scalar-tensor theories discussed in \tsec{sec:linearperturbations} for these extrapolations instead, the thresholds differ in general, $\nu\neq\nu_{\rm i}$.
Hence, Eq.~(\ref{eq:extrapolation}) corresponds to defining the peak-threshold at $a_{\rm i}$.

The gravitational force enhancement determined by Eqs.~(\ref{eq:enhf}) and (\ref{eq:thshsphcoll}) and thus, $\deltac$, depend on the environmental density $\denv$ or $\delta_{\rm env,i}$ with larger modifications and stronger suppression for low and high values of $\denv$, respectively.
In Ref.~\cite{li:11a} this environment was specified by its Lagrangian (or initial comoving) size with the environment defined as a spherical region around the same centre as the top-hat overdensity that has a radius that is larger than the halo that will form but small enough to be considered its surrounding.
Refs.~\cite{li:12, li:12b} then characterised the environment in terms of the Eulerian (physical) size, emphasising that the difference in the respective probability distributions of $\denv$ leads to differences in the corresponding structures that are formed with the Eulerian environments being more likely to be larger than their Lagrangian counterparts and hence causing a stronger suppression of the modified gravity effects.
They argued that a scale at the order of the Compton wavelength $\lambda_{\rm C}$ is a natural choice for the Eulerian radius of the environment, setting $\zeta=5h^{-1}~{\rm Mpc}$.
This value has also been adopted by Refs.~\cite{lombriser:13b, lombriser:13c} together with the probability distribution $P_{\zeta}(\denv)$ of the Eulerian environmental density $\denv$ described in Refs.~\cite{lam:08, li:12b} to analyse the environmental dependence of structures formed in chameleon theories, also employing different averaging procedures.
Thereby, using the environmental average of collapse densities, $\langle\deltac\rangle_{\rm env}$, when modelling halo properties (see \tsec{sec:chameleonclusters}) produced a good match to results from $f(R)$ $N$-body simulations, furthermore, providing a simplification over modelling observables with $\deltac(\denv)$ first with subsequent averaging over $P_{\zeta}(\denv)$.
This approach can be further simplified by using the environmental density for which $\deltac$ comes close to this average.
For the cosmology in \tsec{sec:Nbodysimulations} and setting $z_{\rm f}=0$, one can adopt $\langle\deltac\rangle_{\rm env}\approx\deltac(\denv=0.4)$.
Similar approximations such as using the peak of the environmental distribution $\denv\approx0.8$ or the average environment $\langle\denv\rangle_{\rm env}\approx0.16$ for the same cosmological configuration have also been explored in Refs.~\cite{lombriser:13b, lombriser:13c}.
The collapse density $\deltac(\denv=0.4)$ for collapse at $z=0$ and the cosmology defined in \tsec{sec:Nbodysimulations} is shown in Fig.~\ref{fig:sphcoll}.

Note that there are alternative approaches to computing $\deltac$ through the thin-shell force enhancement in the collapse as described in Eqs.~(\ref{eq:enhf}) and (\ref{eq:thshsphcoll}).
Schmidt \etal.~\cite{schmidt:08} studied the spherical collapse in the extreme cases $\Delta F/F_{\rm N}=1/3$ and $\Delta F/F_{\rm N}=0$ in Eq.~(\ref{eq:shellmotion}), corresponding to the full modification and fully screened case in $f(R)$ gravity.
Borisov \etal.~\cite{borisov:11} generalised this approach by considering the isotropic evolution of an initial overdensity profile according to Eqs.~(\ref{eq:fRsim}) and (\ref{eq:potsim}), which in this case becomes a one-dimensional problem.
They find that, whereas in the limiting cases studied by Ref.~\cite{schmidt:08}, an initial top-hat overdensity remains a top hat at later times, under this evolution, the initial top-hat overdensity develops a spike at its edge, indicating shell crossing.
To evade the consequent numerical problems, they introduce a Gaussian smoothing of the initial profile to its cosmological background with its dispersion as a free parameter.
However, due to the breakdown of Birkhoff's theorem in $f(R)$ gravity and the associated environmental dependence, the collapse density becomes dependent on the choice of initial profile and, hence, the dispersion parameter.
Kopp \etal.~\cite{kopp:13} followed this approach, introducing a method based on peaks theory and the matter transfer function that, given the cosmological parameters, fixes the initial overdensity profile. 
Moreover, whereas Ref.~\cite{borisov:11} use the scale and time dependent linear growth function of $f(R)$ gravity in Fourier space, convolved with the Fourier image of a top hat, to define $\deltac$, Ref.~\cite{kopp:13} define the effective $\deltac$ evolved from the initial overdensity according to the linear $\Lambda$CDM growth function $D(a)$ as in Eq.~(\ref{eq:extrapolation}).
Ref.~\cite{kopp:13} give a fitting function for their $\deltac$ as a function of redshift, halo mass, $\Om$, and $\absfR$.
The corresponding limit for the spherical collapse density in the cosmology specified in \tsec{sec:Nbodysimulations} at $z_{\rm f}=0$ is shown in Fig.~\ref{fig:sphcoll}.
Finally, the chameleon transition in the peak threshold $\nu$ can also be modelled via the phenomenological approach of Li and Hu~\cite{li:11b}, which will be discussed in more detail in \tsec{sec:halomassfunctionlinearhalobias}, where the effective collapse density can be reconstructed from $\deltac \equiv \deltacLCDM\sqrt{S/S_{\rm PPF}}$ with $\deltacLCDM$ being the $\Lambda$CDM spherical collapse density and $S_{\rm PPF}$ is given by Eq.~(\ref{eq:PPF}).


\subsection{Chameleon clusters} \label{sec:chameleonclusters}


Galaxy clusters are of particular interest when searching for cosmological signatures of a chameleon field and placing constraints on the models.
The enhanced growth of structure due to the extra force exerted by the scalar field yields, for instance, an increase in the abundance of massive clusters, as discussed in \tsec{sec:halomassfunctionlinearhalobias}.
Counteracting this modification is a decrease of the growth enhancement near the Compton wavelength of the background scalar field, beyond which gravity returns to Newtonian, causing a decrease of the overabundance of massive halos for decreasing values of $1-\bscal$.
More importantly, however, the chameleon mechanism yields a more effective, nonlinear recovery of the Newtonian results, which is not just dependent on the mass of the halo but also on its environment.
Similarly, with the increased abundance of massive halos, they become less biased, where the chameleon mechanism acts again to recover the Newtonian results.

Equally interesting are the chameleon field and matter density profiles within the cluster.
The gravitational modifications yield an increase in halo concentrations for halo masses defined at equal overdensities, accordingly, with enhanced characteristic matter densities and furthermore, a matter pile-up in the infall region of the clusters.
While for values of $\scal\simeq1$, light deflection by this matter distribution is equivalent to the general relativistic effect, the dynamically inferred matter distributions differ for unshielded clusters and can be determined from the corresponding chameleon field profile as discussed in \tsecs{sec:clusterdensityprofile}, \ref{sec:chameleonfieldprofile}, and \ref{sec:dynamicalmassprofile}.


\subsubsection{Halo mass function and linear halo bias} \label{sec:halomassfunctionlinearhalobias}



\begin{figure*}
 \centering
 \resizebox{\hsize}{!}{
  \resizebox{0.5\hsize}{!}{\includegraphics{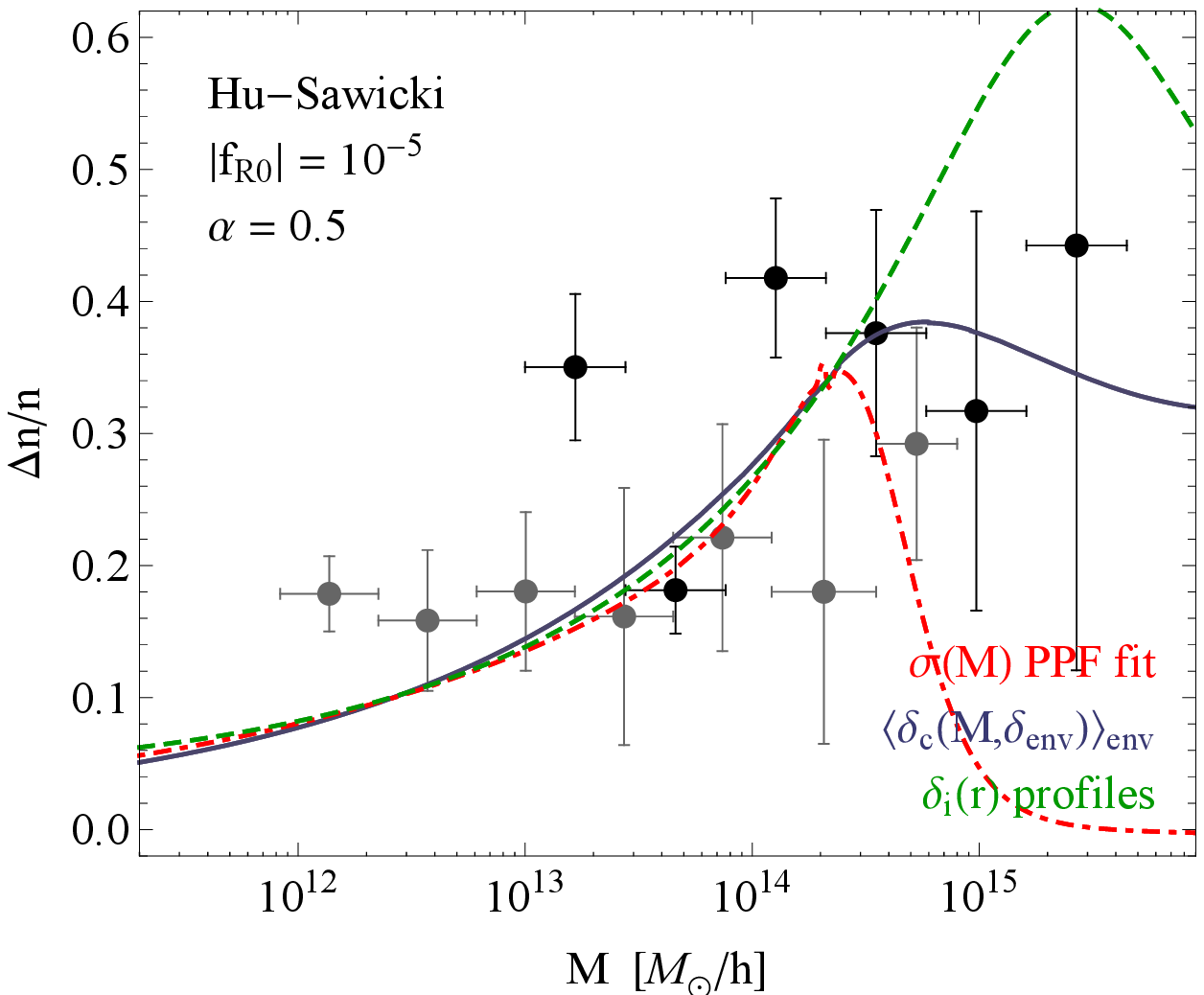}}
  \resizebox{0.5\hsize}{!}{\includegraphics{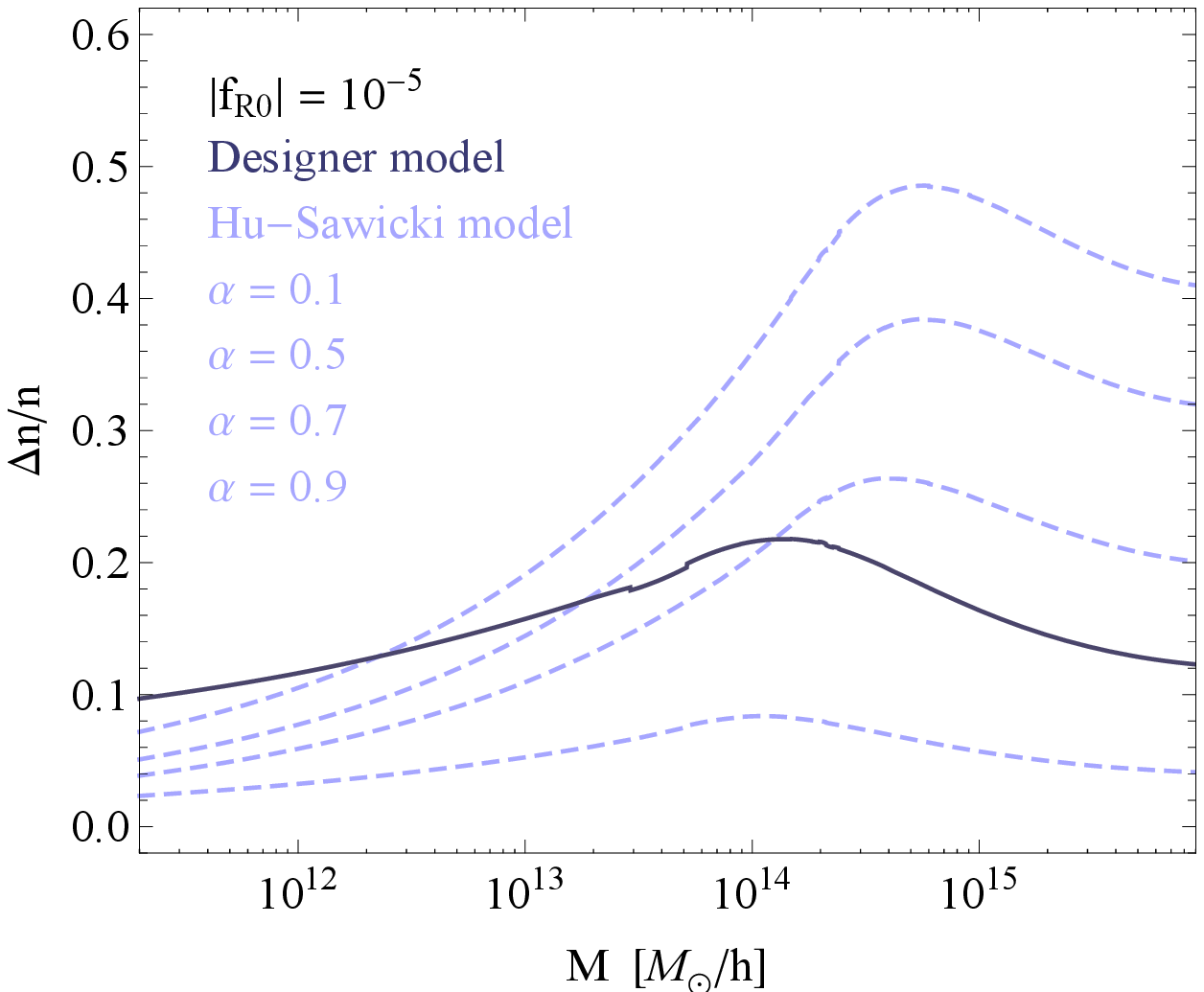}}
 }
 \caption{
  \emph{Left panel:} Relative enhancement in the halo mass function at $z=0$ of the Hu-Sawicki $\absfR=10^{-5}$ model ($\alpha=0.5$) with respect to $\Lambda$CDM using the different spherical collapse densities shown in Fig.~\ref{fig:sphcoll} in the Sheth-Tormen prescription, Eq.~(\ref{eq:st}).
  Also shown are the $N$-body simulations of Ref.~\cite{li:12c} (black data points) and Ref.~\cite{zhao:10b} (gray data points).
  \emph{Right panel:} Comparison between the corresponding enhancements in the designer model and the Hu-Sawicki model with different values of $\alpha$.  
  }
\label{fig:hmf}
\end{figure*}


The statistics of virialised clusters can be described using excursion set theory, where collapsed structures are associated with regions where the initial matter density fields smoothed over these regions exceed the threshold $\deltac$.
The variance $S$ in Eq.~(\ref{eq:variance}) characterises the size of such a region and when varied causes incremental steps in the smoothed initial overdensity field.
These steps are independent of their previous values if the wavenumbers are uncorrelated, describing a Brownian motion of the smoothed matter density field with Gaussian probability distribution and with the increment being a Gaussian field with zero mean.
The distribution $f$ of the Brownian motion trajectories which cross the flat barrier $\deltac$ the first time at $S$ was characterised by Press and Schechter~\cite{press:74}.

In the case of chameleon gravity, the barrier is, however, no longer flat and depends on both the variance and the environment that embeds the collapsing halo.
Even in the Newtonian case, the barrier becomes dependent on the variance once relaxing the assumption of sphericity of the halo.
Sheth and Tormen~\cite{sheth:99} modified the Press-Schechter first-crossing distribution based on excursion set results with the moving barrier of ellipsoidal collapse~\cite{sheth:99b, sheth:01},
\bq
 \nu \, f(\nu) = \mathcal{N} \sqrt{\frac{2}{\pi} q \, \nu^2} \left[ 1 + \left(q \, \nu^2\right)^{-p} \right] e^{-q\,\nu^2/2} \label{eq:st}
\eq
with peak-threshold $\nu\equiv\deltac/\sqrt{S}$, normalisation $\mathcal{N}$ such that $\int \rmd \nu \, f(\nu) = 1$, as well as $p=0.3$ and $q=0.707$.
The parameter $q$ is set to match the halo mass function 
\bq
 n_{\ln\Mvir} \equiv \frac{\rmd n}{\rmd\ln\Mvir} = \frac{\rhomb}{\Mvir} f(\nu) \frac{\rmd \nu}{\rmd\ln\Mvir} \label{eq:hmf}
\eq
with measurements from $\Lambda$CDM $N$-body simulations.
Although Eq.~(\ref{eq:st}) has been obtained in the concordance model, in combination with the mass and environment dependent spherical collapse model in \tsec{sec:sphericalcollapsemodel}, it also provides a good description of the relative enhancement of the chameleon halo mass function with respect to its $\Lambda$CDM counterpart using the same values of $p$ and $q$~\cite{lombriser:13b,lombriser:13c}.
The corresponding enhancements for the Hu-Sawicki and designer $f(R)$ models are shown in Fig.~\ref{fig:hmf}.
Also shown are the enhancements measured in the Hu-Sawicki $\alpha=0.5$ $N$-body simulations of Ref.~\cite{li:12c} (black data points) and Ref.~\cite{zhao:10b} (gray data points).
In comparison, the designer model yields a smaller increase in the halo mass function for massive halos, whereas the enhancement is larger for small masses.
This behaviour reflects the differences in $\deltac$ and $f_R/f_{R0}$ between the two models illustrated in Figs.~\ref{fig:constraints} and \ref{fig:sphcoll}.

Instead of using the phenomenological halo mass function in Eq.~(\ref{eq:st}), one can use excursion set theory to compute a theoretically better motivated first-crossing distribution with the moving barrier defined by the linear chameleon collapse density $\deltac(S,\denv)$.
The corresponding halo mass function is then determined from Eq.~(\ref{eq:hmf}).
This approach was conducted in Ref.~\cite{li:11a} using a Lagrangian definition of environment and extended to Eulerian environments in Ref.~\cite{li:12b}, who compared the two approaches as well as performed both numerical integrations and Monte Carlo simulations, finding that due to the larger likelihood of high-density environments in the Eulerian case, the overabundance of medium and large size halos with respect to $\Lambda$CDM is weakened.
The halo mass function of the Hu-Sawicki $f(R)$ model using excursion set theory and a numerical integration method has been computed in Ref.~\cite{lombriser:13b}.
The comparison to $N$-body simulations showed a better agreement with the Sheth-Tormen approach when combined with the mass and environment dependent spherical collapse model and a subsequent averaging over the probability distribution of the Eulerian environment.
Ref.~\cite{kopp:13} formulated an analytic expression for the halo mass function based on excursion set theory with a drifting and diffusing barrier computed from the $f(R)$ evolution of the initial density profile with uncorrelated steps, which they tested against Monte Carlo random walks.
Note that in Fig.~\ref{fig:hmf}, the results referring to the smoothed initial density profile of Ref.~\cite{kopp:13} are not using this result for the halo mass function but apply the spherical collapse density from this approach (see \tsec{sec:sphericalcollapsemodel}) to the Sheth-Tormen formula in Eq.~(\ref{eq:st}).

Finally, whereas in all of these approaches, the random walk was considered Markovian, i.e., with uncorrelated steps, Lam and Li~\cite{lam:12b} introduce correlated steps in the excursion set approach and find that, in general, this leads to an enhancement of the modifications in the halo mass function.
The difference with respect to uncorrelated steps is due to a change of distribution of the Eulerian environmental densities, whereas the Lagrangian definition is not affected, and hence, the first-crossing probability, as well as correlations between $\delta$ and $\denv$.
In order to analyse these effects in the first-crossing distribution, they study three different window functions, a sharp $k$-filter for the uncorrelated case, as well as a Gaussian and top-hat filter for the correlated steps, for which they run Monte Carlo simulations.

Alternatively to introducing the chameleon mechanism in the spherical collapse density, Li and Hu~\cite{li:11b} proposed a nonlinear parametrised post-Friedmann (PPF) description to determine the halo mass function for chameleon $f(R)$ gravity.
They phenomenologically interpolate between the linearised and suppressed regimes by introducing a chameleon PPF transition in the variance as
\bq
 S_{\rm PPF}^{1/2}(M) = \frac{S_{\scal}^{1/2}(M) + \left(M/M_{\rm th}\right)^{\mu} S_{\Lambda{\rm CDM}}^{1/2}(M)}{1+\left(M/M_{\rm th}\right)^{\mu}}, \label{eq:PPF}
\eq
where $M_{\rm th}$ and $\mu$ are fitted simultaneously to different halo mass functions extracted from $N$-body simulations with different configurations of the chameleon field.
The PPF peak threshold in Ref.~\cite{li:11b} is then given by
\bq
 \nu_{\rm PPF} \equiv \frac{\deltacLCDM}{S_{\rm PPF}^{1/2}(M)},
\eq
which they subsequently use in the Sheth-Tormen expression Eq.~(\ref{eq:st}) to approximate the halo mass function.
The parameters $(M_{\rm th},\mu)$ have been calibrated to $N$-body simulations of the Hu-Sawicki $f(R)$ gravity model in Refs.~\cite{li:11b,lombriser:13b}.
With the mass definition $M_{200}$ from setting $\Delta=200$, Ref.~\cite{li:11b} finds $M_{\rm th} = 1.345\times10^{13} \left( 10^6 \absfR \right)^{3/2}~\Msunh$ and $\mu=2.448$.
The corresponding PPF enhancement in the halo mass function is shown in Fig.~\ref{fig:hmf}, where masses have been rescaled to $M_{390}$ following Ref.~\cite{lombriser:13b}.
Note that if attributing the chameleon transition in the peak threshold $\nu_{\rm PPF}$ and accordingly in the halo mass function to a modification of the spherical collapse density rather than to a transition in the variance (see \tsec{sec:sphericalcollapsemodel}), the scaling of $M_{\rm th}$ with $\absfR^{3/2}$ can be derived from the spherical collapse model when considering the coefficient of the force modification in Eq.~(\ref{eq:thshsphcoll}) that scales as $(1-\scal)/r_{\rm th}^2\sim(1-\scal)/M^{2/3}$.


Finally, with the Sheth-Tormen halo mass function Eq.~(\ref{eq:st}), the linear halo bias obtained in the peak-background split becomes~\cite{sheth:99}
\bq
 b_{\rm L}(\Mvir) \equiv b(k=0,\Mvir) = 1 + \frac{a\,\nu^2-1}{\deltac} + \frac{2p}{ \deltac \left[ 1 + \left( a\,\nu^2 \right)^p \right] }. \label{eq:blin}
\eq
The effective linear collapse density $\deltac$ in chameleon models is suppressed relative to $\Lambda$CDM, which causes $b_{\rm L}$ to decrease.
Moreover, the modification becomes mass and environment dependent and can be determined using the spherical collapse model described in \tsec{sec:sphericalcollapsemodel}.

Ref.~\cite{schmidt:08} performed a measurement of the linear halo bias in $N$-body simulations of Hu-Sawicki $f(R)$ gravity and found good agreement with the deviations in $b_{\rm L}$ predicted by Eq.~(\ref{eq:blin}) when using the modified peak-threshold.
Halo biasing has also been analysed for chameleon models in Ref.~\cite{lam:12b} within the framework of excursion set theory using the unconditional and conditional first-crossing distribution with different smoothing window functions.
They find an increase in the modifications of $b_{\rm L}$ for correlated steps with respect to the uncorrelated case due to less likely high-density environments and correlations of $\denv$ with $\delta$ in the non-Markovian scenario. 
Deviations in the halo bias and halo mass function have also been analysed for a Yukawa-type modification of gravity, using the spherical collapse model and excursion set theory based on the scale-dependent modification of the growth function~\cite{hui:07, martino:08, parfrey:11}.


\subsubsection{Cluster density profile} \label{sec:clusterdensityprofile}


Navarro, Frenk, and White (NFW)~\cite{navarro:95} found that the dark matter clusters formed in $\Lambda$CDM $N$-body simulations are well described by spherical halos with the simple universal radial density profile
\bq
 \delta\rhom(r) = \frac{\rho_{\rm s}}{\frac{r}{r_{\rm s}} \left( 1+\frac{r}{r_{\rm s}} \right)^2 }. \label{eq:nfw}
\eq
Hereby, $\rhos$ and $\rs$ denote the characteristic density and scale, respectively, which can be calibrated to simulations.
Alternatively, given a specific virial halo mass $\Mvir$ that is defined by the virial overdensity $\Delta_{\rm vir}$, one may use the virial halo concentration $\cvir\equiv\rvir/\rs$ as the free parameter and the relations
\bqa
 \rhos & = & \frac{1}{3} \bar{\rho}_{\rm m} \Delta_{\rm vir} \cvir^3 \left[ \ln(1+\cvir) - \frac{\cvir}{\cvir+1} \right]^{-1}, \label{eq:chardens} \\
 \rs & = & \frac{1}{\cvir}\left( \frac{3\Mvir}{4\pi\bar{\rho}_{\rm m}\Delta_{\rm vir}} \right)^{1/3}, \label{eq:charscale}
\eqa
in the density profile Eq.~(\ref{eq:nfw}).
The NFW profile can further be reduced to a function of halo mass only by adopting a mass-concentration scaling relation.
In the following, the concentration shall be given by the relation
\bq
 \cvir(\Mvir,a) = 9 a \left(\frac{\Mvir}{M_*}\right)^{-0.13}
 \label{eq:cMlcdm}
\eq
with the critical mass $M_*$ satisfying $S(M_*)=\deltac^2$.
Eq.~(\ref{eq:cMlcdm}) has been calibrated to approximately $5\times10^3$ halos of mass $10^{11}-10^{14}~\Msunh$ extracted from $\Lambda$CDM $N$-body simulations in Ref.~\cite{bullock:99}
and shall here be assumed to also apply to more massive halos, concentrating on $\Delta_{\rm vir}\equiv390$ as in \tsec{sec:Nbodysimulations}.

When gravity is modified, Eqs.~(\ref{eq:nfw}) and (\ref{eq:cMlcdm}) may not be applicable and it is worth checking their performance against $N$-body simulations.
Ref.~\cite{lombriser:12} found that Eq.~(\ref{eq:nfw}) provides comparably good fits to the dark matter halo density profiles extracted from $N$-body simulations of the Hu-Sawicki $f(R)$ model as to the concordance model halos.
Given that the fit is accurate for a range of $\absfR$ values and, accordingly, for different magnitudes of the gravitational modifications and the similarity of the chameleon mechanism between different choices of model parameters in Eq.~(\ref{eq:thshsphcoll}), this motivates the use of the NFW profiles for a wider class of chameleon models defined by Eqs.~(\ref{eq:horndeski}), (\ref{eq:jordanbransdicke}), and (\ref{eq:potential}).
The scaling relation Eq.~(\ref{eq:cMlcdm}) can be adopted in the chameleon models with the replacement $S \rightarrow S_{\scal}$, where $S_{\scal}$ is the variance computed using the scale-dependent growth function $D_{\scal}(a,k)$~\cite{schmidt:08,lombriser:11b,li:11b}.
In this approach, however, the chameleon modification is incorporated in $M_*$ in a linear manner, which depends on the amplitude $(\bscal_0-1)$ and only rescales $\cvir(\Mvir)$ with a constant factor.
Hence, it does not capture the chameleon mechanism, neither taking into account dependencies of the modification on mass nor environment~\cite{lombriser:12}.
These dependencies can be introduced in $\cvir$ by reinterpreting the concentration-mass relation defining the critical mass as $M_*(\deltac,\sigma)\equiv\sigma^{-1}\circ\deltac$ and adopting the effective $\deltac$ from the chameleon spherical collapse model~\cite{lombriser:13c}.
The chameleon mechanism is then incorporated in the concentration via the relations
\bqa
 \cvir(\Mvir,\denv,a) & = & 9 a \left[ \frac{M_*(\Mvir,\denv)}{\Mvir} \right]^{0.13}, \label{eq:concentration} \\
 M_*(\Mvir,\denv) & \equiv & (\sigma^{-1}\circ\deltac)(\Mvir,\denv) \nonumber\\
 & = & \sigma^{-1}(\deltac(\Mvir,\denv)). \label{eq:massstar}
\eqa
As $\deltac$ becomes smaller in the chameleon models, $M_*$ and the concentration become larger, causing an enhancement in $\rhos$ and a decrease of $\rs$ compared to their $\Lambda$CDM counterparts.
This behaviour is in agreement with the measurements of halo concentration, characteristic density, and characteristic radius from $N$-body simulations of the Hu-Sawicki $f(R)$ model~\cite{lombriser:12}.

Finally, while within the virial radius, the halo density profiles of the chameleon and concordance model clusters agree up to different values of characteristic density and radius, at a few virial radii, $N$-body simulations show an enhancement of the halo-matter correlation function due to a matter pile-up in the infall region caused by the late-time enhanced gravitational forces~\cite{schmidt:08, lombriser:11b} (also see~\cite{zu:13}).
The effect can be well described by the halo model~\cite{lombriser:11b}.


\subsubsection{Chameleon field profile} \label{sec:chameleonfieldprofile}


Given the dark matter profile of a cluster discussed in \tsec{sec:clusterdensityprofile}, i.e., the NFW relation for $\drhom$ in Eq.~(\ref{eq:nfw}), the scalar field profile within the halo can be obtained from solving the scalar field equation, Eq.~(\ref{eq:sfeq}).
Assuming sphericity and the quasistatic limit, this yields a second-order differential equation for $\scal(r)$, which can easily be integrated numerically adopting the substitution $\scal-1 = - e^{u(r)}$~\cite{oyaizu:08a, schmidt:10, zhao:10b, lombriser:12}.

Alternatively, an analytic approximation for the chameleon field profile can be derived requiring $\scal\simeq1$ and linearising the scalar field potential $U(\scal)$ in Eq.~(\ref{eq:sfeq}) with respect to $\bscal$.
After subtraction of the background, $\dscal \equiv \scal - \bscal$, the scalar field equation becomes
\bq
 \tilde{\nabla}^2 \dscallin - \mcom^2 \dscallin + \frac{\kappa^2}{3+2\omega} \delta\rhom \simeq 0, \label{eq:qssfeq}
\eq
which is solved by~\cite{lombriser:12}
\bqa
 \dscallin & \simeq & -\frac{\kappa^2\rhos\rs^3}{6+4\omega} \left\{ \Gamma[0,\mcom(r+\rs)] e^{2\mcom(r+\rs)} \right. \nonumber\\
 & & + \Gamma[0,-\mcom(r+\rs)] - \Gamma(0,-\mcom\,\rs) \nonumber\\
 & & \left. - e^{2\mcom\,\rs}\Gamma(0,\mcom\,\rs) \right\} \frac{e^{-\mcom(r+\rs)}}{r} \label{eq:dscallin}
\eqa
with the upper incomplete gamma function
\bq
 \Gamma(s,r) = \int_r^{\infty} \rmd t \, t^{s-1} e^{-t}.
\eq
The integration constants are set by the requirements that $\lim_{r\rightarrow\infty}\dscallin=0$ and $\lim_{r\rightarrow0}r\dscallin=0$.
In this approximation, the chameleon transition is assumed to take place instantaneously once $\dscal=1-\bscal$, hence,
\bq
 \dscal \approx \min \left( \dscallin, 1-\bscal \right) \label{eq:instrans}
\eq
or equivalently, $\scal \approx \min \left(\scal_{\rm lin}, 1 \right)$.
For $\rhom\gg\rhomb$, the linearised scalar field simplifies to~\cite{lombriser:12}
\bq
 \dscallin \simeq \frac{\kappa^2 \rho_{\rm s} r_{\rm s}^3}{3+2\omega} \left[ \frac{\ln (1+r/r_{\rm s})}{r} - \mcom\,e^{\mcom\,r_{\rm s}} \Gamma(0,\mcom\,r_{\rm s}) \right], \label{eq:hdlinsf}
\eq
for which the chameleon screening scale is~\cite{lombriser:13c}
\bq
 \rcham = -r_{\rm s} - A^{-1} W\left[-A\,r_{\rm s} \exp(-A\,r_{\rm s})\right]
\eq
with the Lambert $W$ function $W[\cdot]$, solving $x=W(x)\exp[W(x)]$, and
\bq
 A \equiv \frac{3+2\omega}{\kappa^2 \rho_{\rm s} r_{\rm s}^3}(1-\bscal) + \mcom\, e^{\mcom\, r_{\rm s}} \Gamma(0,\mcom\,r_{\rm s}).
\eq

Rather than assuming an instantaneous transition of the linearised scalar field into a chameleon-shielded regime, one may want to require a continuously differentiable transition in $\dscal$~\cite{pourhasan:11}.
In this case, a free $\rcham$ and the two integration constants of the outer linearised solution $\dscal_{\rm out}$ obtained from the integration of Eq.~(\ref{eq:qssfeq}) in the limit of $\rhom\gg\rhomb$, i.e., neglecting the term $-\mcom^2\dscal$, are matched to the inner chameleon-shielded solution $\dscal_{\rm in}$, where $U_{\scal} \simeq R/2$, which for the potential Eq.~(\ref{eq:potential}) takes the expression Eq.~(\ref{eq:scalfield}).
More specifically, the chameleon field
\bq
 \dscal = \left\{
 \begin{array}{ll}
 \dscal_{\rm out}, & r>\rcham \\
 \dscal_{\rm in}, & r\leq\rcham
 \end{array}
 \right.
 \label{eq:dscalC1}
\eq
and $\rcham$ are required to satisfy the conditions
\bqa
  \dscal_{\rm out}(\rcham) & = & \dscal_{\rm in}(\rcham), \label{eq:dscalcond} \\
  \dscal'_{\rm out}(\rcham)& = & \dscal'_{\rm in}(\rcham), \label{eq:dscalpcond}
\eqa
and that $\lim_{r\rightarrow\infty}\scal_{\rm out}(r)=\scal_{\rm env}$, where $\scal_{\rm env}\simeq\bscal$ will be assumed.
While in the resulting relations, one integration constant is set by the environment, the other is set by the transition scale $\rcham$, which has to be computed numerically.

The different profiles discussed here have been compared to each other in Ref.~\cite{lombriser:12} showing good agreement with $N$-body simulations of the Hu-Sawicki $f(R)$ model.
Note, however, that in all of these approaches, requiring that $\lim_{r\rightarrow\infty}\dscal_{\rm out}(r)=0$ does not necessarily yield a recovery of the simulated chameleon field profile of a particular cluster as the environment of the cluster can deviate from the cosmological background.
In this case, $\dscal$ should be matched to boundary conditions such as obtained from extracting $\dscal(\rvir)$ from simulations~\cite{lombriser:12}.
Furthermore, note that the scalar field profile obtained from Eqs.~(\ref{eq:dscalC1}), (\ref{eq:dscalcond}), and (\ref{eq:dscalpcond}) does not apply on scales approaching the Compton wavelength, where for scales beyond it the scalar field is suppressed according to Eq.~(\ref{eq:dscallin}).

Following Refs.~\cite{terukina:12, terukina:13}, instead of solving $\rcham$ numerically in the last approach,
given that $\rhom\gg\rhomb$, one can approximate the inner solution in Eq.~(\ref{eq:dscalC1}) as $\scal_{\rm in}\simeq0$, which yields
\bqa
 \rcham & \simeq & \frac{\kappa^2\rhos\rs^3}{3+2\omega}\frac{1}{1-\bscal}-\rs, \label{eq:rchamapp} \\
 \dscal_{\rm out} & \simeq & \frac{\kappa^2\rhos\rs^3}{3+2\omega}\ln\left(\frac{r+\rs}{\rcham+\rs}\right)\frac{1}{r} + (1-\bscal)\frac{\rcham}{r}. \label{eq:dscaloutapp}
\eqa

%


\subsubsection{Dynamical mass profile} \label{sec:dynamicalmassprofile}


An unscreened test particle of mass $m_{\rm t}$ in the cluster experiences the extra force
\bq
 F_{\scal} \equiv -\frac{m_{\rm t}}{2} \tilde{\nabla} \scal.
\eq
due to the presence of the chameleon field.
This extra force can be interpreted as being exerted by the phantom mass
\bq
 M_{\scal} \equiv -\frac{r^2}{2G} \frac{\rmd}{\rmd r} \scal(r),
 \label{eq:scalarfieldmass}
\eq
assuming a spherical system and $\scal\simeq1$.
Eq.~(\ref{eq:scalarfieldmass}) contributes to the dynamically inferred mass as $M_{\rm D}=M+M_{\scal}$.
Using Eqs.~(\ref{eq:dscalC1}), (\ref{eq:rchamapp}), and (\ref{eq:dscaloutapp}), it therefore follows that
\bq
 M_{\rm D}(r) \simeq \left\{ 1 + \frac{\Theta(r-\rcham)}{3+2\omega}\left[ 1 - \frac{M(\rcham)}{M(r)} \right] \right\} M(r), \label{eq:dynamicalmass}
\eq
where $\Theta$ is the Heaviside step function.
The cluster mass profile
\bq
 M(r) = 4\pi\rhos\rs^3 \left[ \ln\left(1+\frac{r}{\rs}\right) - \frac{r}{r+\rs} \right] + M_{\rm c}
\eq
is obtained from integrating the NFW density profile, where $M_{\rm c}$ is a mass correction due to deviations from this relation in the inner part of the cluster.
The approximations and equations leading to the expression Eq.~(\ref{eq:dynamicalmass}) generalise and yield a derivation of the assumption made for $M_{\rm D}(r)$ in Ref.~\cite{schmidt:10}.
Note, however, that Eq.~(\ref{eq:dynamicalmass}) does not apply on scales approaching the Compton wavelength, where for scales beyond it the scalar field is suppressed according to Eq.~(\ref{eq:dscallin}) (see \tsecs{sec:linearperturbations} and \ref{sec:chameleonfieldprofile}).
The difference between $M_D(r)$ and $M(r)$ also depends on the environment, which in Eq.~(\ref{eq:dynamicalmass}) can be taken into account through dropping the condition $\scal_{\rm env}=\bscal$.
It has also been shown in Ref.~\cite{li:12} that at $r=\rvir$, the difference in mass observed in $N$-body simulations of $f(R)$ gravity can be well described using the thin-shell condition.

Moreover, dense objects such as stars may not feel the full force modification in Eq.~(\ref{eq:dynamicalmass}) due to self-shielding in the chameleon mechanism.
The effective force felt by such an object can be estimated by Eqs.~(\ref{eq:thinshell}) and (\ref{eq:enhf}), where $\scal_{\rm out}$ is determined by its environmental field value.
While the stars may be screened or partially screened, the gas of a cluster feels the full force modification.
Assuming hydrostatic equilibrium of the gas, for a spherically symmetric system, the gas density $\rho_{\rm gas}$ and pressure $P$ relate to the dynamical mass profile as
\bq
 \frac{1}{\rho_{\rm gas}(r)}\frac{\rmd P(r)}{\rmd r} = - \frac{G\,M_{\rm D}(r)}{r^2}.
\eq
Assuming no contribution from non-thermal pressure, the gas pressure relates to the gas density and temperature as $P=P_{\rm thermal}\propto\rho_{\rm gas}T_{\rm gas}$.
Whereas the mass profile $M(r)=M_{\rm D}(r)-M_{\scal}(r)$ can be measured by weak gravitational lensing around a cluster, the gas temperature, density, and pressure can be determined from X-ray observations and the Sunyaev-Zel'dovich effect.
In hydrostatic equilibrium, these observables are uniquely determined from any combination of two of these profiles.
Hence, a combination of these measurements yields a powerful test of gravity~\cite{terukina:13}.


\subsection{Matter power spectrum} \label{sec:matterpowerspectrum}

\begin{figure*}
 \centering
 \resizebox{\hsize}{!}{
  \resizebox{0.5\hsize}{!}{\includegraphics{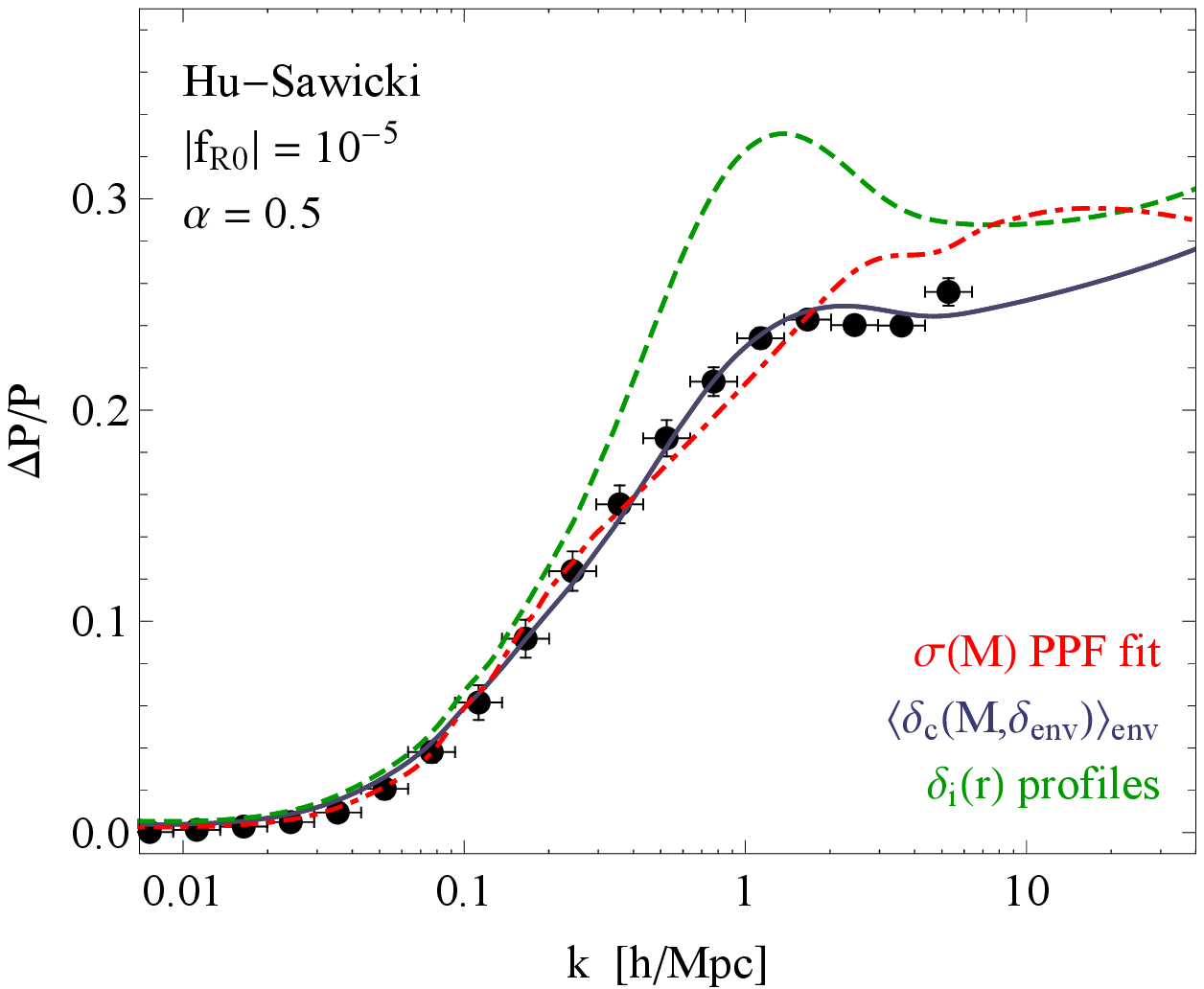}}
  \resizebox{0.5\hsize}{!}{\includegraphics{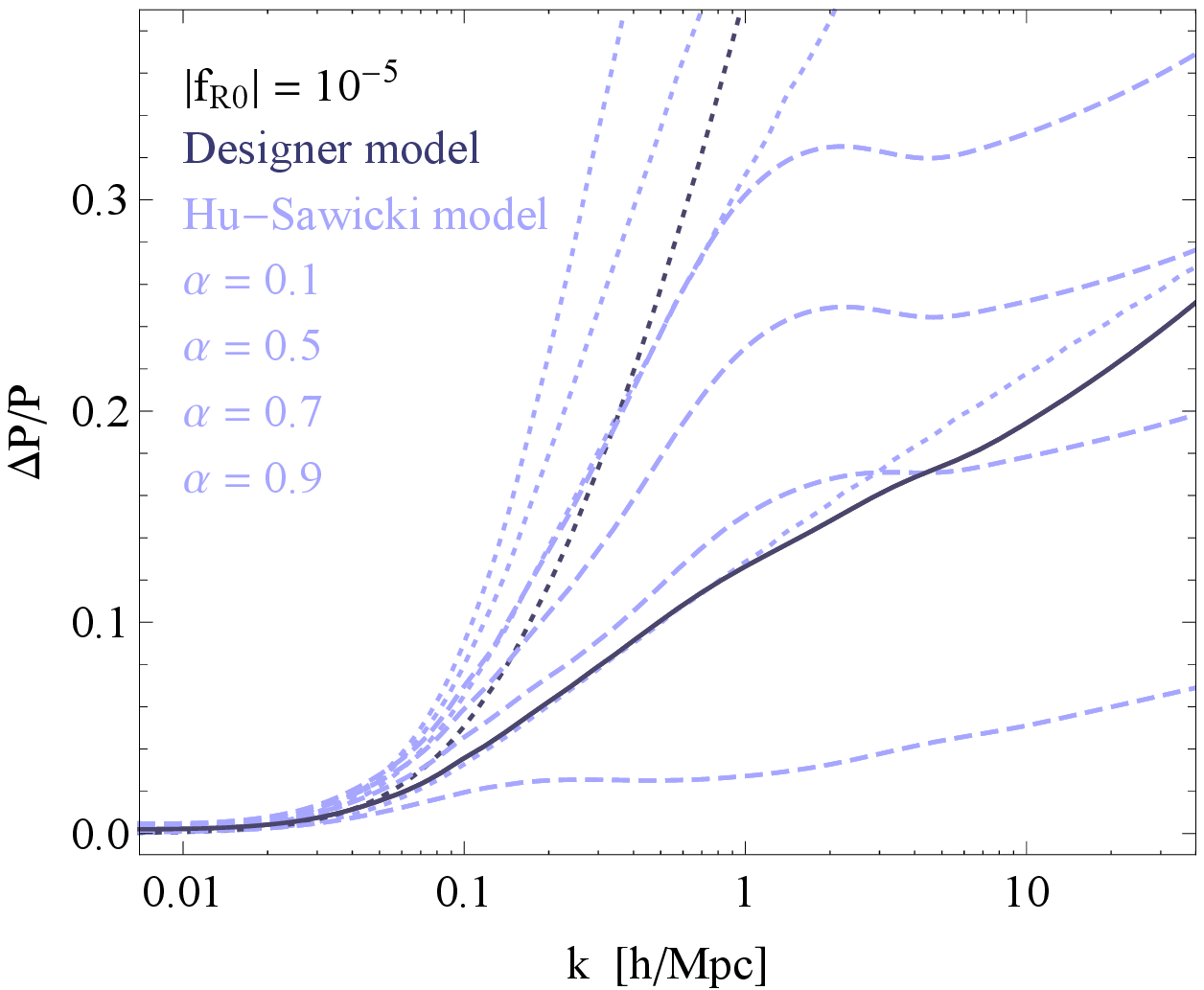}}
 }
 \caption{
  \emph{Left panel:} Relative enhancement of the matter power spectrum at $z=0$ in the Hu-Sawicki model for $\alpha=0.5$ and $\absfR=10^{-5}$ with respect to $\Lambda$CDM predicted by halo model approaches with the different spherical collapse densities of Fig.~\ref{fig:sphcoll}.
  The data points represent results from $N$-body simulations of Ref.~\cite{li:12c}.
  \emph{Right panel:} Comparison between the matter power spectrum enhancements at $z=0$ in the designer model (solid curve) and the Hu-Sawicki model (dashed curves) with different values of $\alpha$ using the halo model and the mass and environment dependent spherical collapse model.
  Dotted curves indicate linear results.
  }
\label{fig:matterpower1}
\end{figure*}

In the halo model~\cite{peacock:00, seljak:00, cooray:02}, statistics of cosmological structures are decomposed into the underlying halo contributions.
The nonlinear matter power spectrum is described by the two-halo and one-halo terms,
\bqa
 P_{\rm mm}(k) & \simeq & P^{2h}(k) + P^{1h}(k), \label{eq:nlpmm} \\
 P^{1h}(k) & = & \int \rmd\ln\Mvir n_{\ln\Mvir} \frac{\Mvir^2}{\rhomb^2} \left| y(k,\Mvir) \right|^2, \label{eq:p1hk}
\eqa
where $y(k,M)$ denotes the Fourier transform of the halo density profile, which shall be given by the NFW expression Eq.~(\ref{eq:nfw}), with a truncation at $\rvir$ and normalisation $\lim_{k\rightarrow0} y(k,M) = 1$.
The two-halo term shall be approximated by $P^{2h}(k)\approx P_{\rm L}(k)$, and the halo mass function and halo concentration shall be determined as described in \tsecs{sec:halomassfunctionlinearhalobias} and {\ref{sec:clusterdensityprofile}}.
Note that the use of the linear matter power spectrum as the two-halo contribution underestimates nonlinear effects on the transition scales between the one-halo and two-halo terms.
Although this is also a problem for modelling the $\Lambda$CDM power spectrum, for chameleon models, the situation is further complicated as the suppression mechanism is not incorporated in the linearly computed growth enhancement.
In Ref.~\cite{lombriser:13c}, it was shown that the relative enhancement in the chameleon matter power spectrum with respect to its $\Lambda$CDM counterpart can be well described by introducing a simple scale-dependent correction to the linear power spectrum
\bq
 P_{{\rm L}\scal}^{\rm eff}(a,k) = \frac{P_{{\rm L}\scal}(a,k) + (k/k_*)P_{{\rm L}\Lambda{\rm CDM}}(a,k)}{1+k/k_*}, \label{eq:transitioncorrection}
\eq
where $k_*\approx0.1\sqrt{(1-\bscal)/10^{-5}}~\Mpch$.
The scale dependence in this correction can be motivated by the relation between the top-hat size and scalar field amplitude in the coefficient of the force modification used in the spherical collapse computation in Eq.~(\ref{eq:thshsphcoll}).
The resulting enhancements of the matter power spectrum for the Hu-Sawicki and designer $f(R)$ models are shown in Fig.~\ref{fig:matterpower1}, where initial power spectra are determined using the Eisenstein-Hu transfer function~\cite{eisenstein:97a,eisenstein:97b}.
The halo mass function and concentration in the one-halo term Eq.~(\ref{eq:p1hk}) are computed using the different spherical collapse densities described in \tsec{sec:sphericalcollapsemodel}.
In the case of the mass and environment dependent spherical collapse model, the most probable environment is assumed, which corresponds to the environmental density at the peak of its probability distribution.
This can be interpreted as an averaging procedure over the unshielded, shielded, and partially shielded forces acting on the dark matter particles.

A similar approach to the correction of the transition regime in Eq.~(\ref{eq:transitioncorrection}) was conducted in Ref.~\cite{li:11b}, who replace the two-halo term and its interpolation to the one-halo term with the same phenomenology as {\sc halofit}~\cite{smith:02}.
More specifically, $P^{2h}\rightarrow2\pi^2\Delta_{\rm Q}^2(k)/k^3$ with
\bq
 \frac{\Delta_{\rm Q}^2(k)}{\Delta_{\rm L}^2(k)} = \frac{\left[1+\Delta_{\rm L}^2(k)\right]^{\beta_n}}{1+\alpha_n\Delta_{\rm L}^2(k)} \exp\left(-\frac{y}{4}-\frac{y^2}{8} \right), \label{eq:DeltaQ}
\eq
where $\Delta_{\rm L}^2(k)= (2\pi^2)^{-1}k^3P_{\rm L}(k)$ and $y=k/k_{\sigma}$ determine the transition scale to the one-halo term with $\sigma_{\rm G}(k_{\sigma}^{-1})=1$ and
\bq
 \sigma_{\rm G}^2(R) = \int \rmd \ln k \, \Delta_{\rm L}^2(k) \exp(-k^2 R^2).
 \label{eq:sigmaG}
\eq
The transition parameters $\alpha_n$ and $\beta_n$ are determined from~\cite{smith:02}
\bqa
 \alpha_n & = & 1.3884 + 0.3700 n_{\rm eff} - 0.1452 n_{\rm eff}^2, \\
 \beta_n & = & 0.8291 + 0.9854 n_{\rm eff} + 0.3401 n_{\rm eff}^2,
\eqa
where
\bq
 n_{\rm eff} \equiv -3 - \left. \frac{\rmd \ln \sigma_{\rm G}^2(R)}{\rmd \ln R} \right|_{\sigma_{\rm G}=1}.
\eq
Predictions of $P(k)$ then vary depending on whether the modified or $\Lambda$CDM linear matter power spectrum is used to determine the transition.
The arithmetic mean between the two power spectra produced by using either $P_{{\rm L}\Lambda{\rm CDM}}$ or $P_{{\rm L}\scal}$ to compute the right-hand side of Eq.~(\ref{eq:DeltaQ}) and $\sigma_{\rm G}$ is shown in Fig.~\ref{fig:matterpower1}.
Note that in this case, the one-halo term is computed according to Ref.~\cite{li:11b}, using the halo mass function and halo concentration determined from the PPF fit in the variance discussed in \tsec{sec:halomassfunctionlinearhalobias}.


\begin{figure}
 \centering
 \resizebox{\hsize}{!}{
  \resizebox{\hsize}{!}{\includegraphics{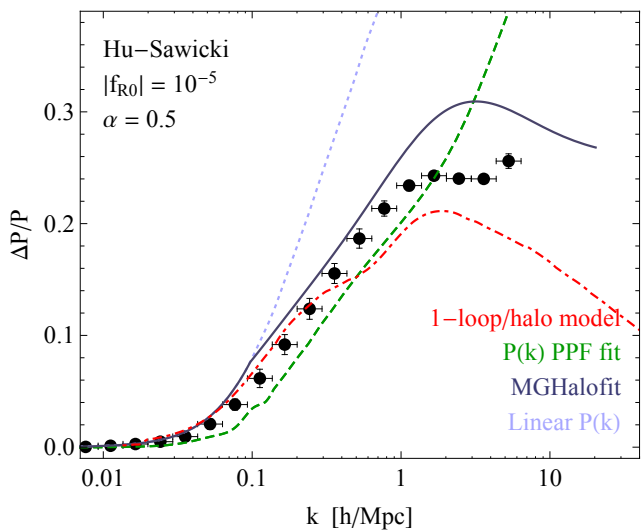}}
 }
 \caption{
  Same as left panel of Fig.~{\ref{fig:matterpower1}} but using {\sc mghalofit} (solid curve), the PPF fit Eq.~(\ref{eq:PkPPFfit}) for $P(k)$ (dashed curve), a combination of one-loop perturbation theory and halo model (dot-dashed curve), and linear perturbations (dotted curve).
  }
\label{fig:matterpower2}
\end{figure}


A different nonlinear PPF formalism, directly modelling $P(k)$, was proposed by Hu and Sawicki~\cite{hu:07b}.
It interpolates between the modified and suppressed regimes of the chameleon matter power spectrum using the nonlinear power spectrum in $\Lambda$CDM $P_{\Lambda{\rm CDM}}$ for the shielded regime and in the unshielded regime, its counterpart $P_{{\rm non}-\Lambda{\rm CDM}}$ that extrapolates the modified linear power spectrum to small scales and does not exhibit a chameleon transition.
The power spectra are then interpolated via the relation
\bq
 P(a,k) = \frac{P_{{\rm non}-\Lambda{\rm CDM}}(a,k) + c_{\rm nl} \Sigma^2(a,k) P_{\Lambda{\rm CDM}}(a,k)}{1 + c_{\rm nl} \Sigma^2(a,k)}. \label{eq:PkPPFfit}
\eq
In Ref.~\cite{hu:07b}, the transition was assumed to scale as $\Sigma^2(k) = (2\pi^2)^{-1}k^3 P_{\rm L}(k)$.
Koyama \etal.~\cite{koyama:09} used perturbation theory and $\Sigma^2(k) = \left[(2\pi^2)^{-1}k^3 P_{\rm L}(k)\right]^{1/3}$ to recover the simulated power spectra in $f(R)$ gravity up to $k\sim0.5\Mpch$.
Zhao et al.~\cite{zhao:10} then extended the PPF approach for $P(k,a)$, fitting
\bq
 \Sigma^2(a,k) = \left[\frac{k^3}{2\pi^2} P_{\rm L}(a,k)\right]^{\alpha_{\rm nl}+\beta_{\rm nl}\,k^{\gamma}},
\eq
to $N$-body simulations of the Hu-Sawicki model with $\alpha=0.5$, where $c_{\rm nl}$, $\alpha_{\rm nl}$, $\beta_{\rm nl}$, and $\gamma$ are re-calibrated at different redshifts and for different values of $\absfR$.
The corresponding matter power spectrum using {\sc halofit} to determine $P_{\Lambda{\rm CDM}}$ and $P_{{\rm non}-\Lambda{\rm CDM}}$ is shown in Fig.~\ref{fig:matterpower2}.
Recently, Zhao~\cite{zhao:13} introduced {\sc mghalofit}, which directly modifies {\sc halofit} by generalising its fitting parameters to accommodate the Hu-Sawicki $f(R)$ model.
The enhancement in $P(k)$ predicted by {\sc mghalofit} is also shown in Fig.~\ref{fig:matterpower2}.

Alternatively, the linear, quasilinear, and nonlinear scales of the matter power spectrum can be computed combining perturbation theory at the one-loop level with the halo model.
An implementation of this concept was developed by Valageas \etal.~\cite{valageas:13} and applied to different scalar-tensor theories, including the Hu-Sawicki $f(R)$ model, by Brax and Valageas~\cite{brax:13b}.
In order to compute the one-loop matter power spectrum, the scalar modification of the Newtonian potential is expanded to third order in the nonlinear density fluctuation.
Subsequently, a Lagrangian-space regularisation is applied to the one-loop expansion and combined with the halo model, using the gravitational modification on the high-mass end of the halo mass function determined by a modified spherical collapse calculation and the NFW density profile with identical mass-concentration relation to the concordance model, hence, not accounting for modified gravity effects.
The results obtained in Ref.~\cite{brax:13b} for the enhancement of the matter power spectrum in the Hu-Sawicki $f(R)$ model with $\alpha=0.5$ are shown in Fig.~\ref{fig:matterpower2}.

Finally, note that in $f(R)$ gravity, nonlinear corrections to the matter power spectrum begin to contribute at slightly larger scales than in $\Lambda$CDM.
The enhancements in the power spectrum at $z=0$ predicted by linear theory can be up to 50\% and 100\% larger at $k=0.1~\Mpch$ and $k=0.2~\Mpch$, respectively, than their nonlinear counterparts.
%


\section{Observational constraints} \label{sec:observationalconstraints}



\begin{table*}
  \begin{center}
  \caption{Selection of current and prospective constraints on $f(R)$ gravity.} \label{tab:constraints}
  \begin{tabular}{lccccc}
    \hline
    \hline
    Measurement & redshift & scale & $\absfR$ constraint & model & Ref. \\
    \hline
    Integrated Sachs-Wolfe (ISW) effect & $\lesssim10$ & $\gtrsim25\hMpc$  & $<3.5\times10^{-1}$ & D    & \cite{song:07, lombriser:10} \\
    Galaxy-ISW cross correlations       & $\lesssim5$  & $\gtrsim40\hMpc$  & $<6.9\times10^{-2}$ & D    & \cite{song:06, giannantonio:09, lombriser:10} \\
    Galaxy power spectrum (WiggleZ)     & $0.2-1$ & $\gtrsim60\hMpc$ & $10^{-4}\lesssim\absfR\lesssim2\times10^{-1}$ & D & \cite{dossett:14} \\
    Galaxy power spectrum (WiggleZ)     & $0.2-1$ & $\gtrsim30\hMpc$ & $\lesssim1.4\times10^{-5}$ & D   & \cite{dossett:14} \\
    Redshift-space distortions (LRG)    & $0.16-0.47$  & $(15-300)\hMpc$   & $\lesssim10^{-4}$   & HS   & \cite{yamamoto:10} \\
    $E_G$ probe                         & $0.32$       & $(10-50)\hMpc$    & $\lesssim10$        & D    & \cite{reyes:10,lombriser:10} \\
    CMB lensing (ACT)                   & $\lesssim6$  & $(1-60)\hMpc$     & $\lesssim10^{-1}$   & D    & \cite{marchini:13a} \\
    CMB lensing (SPT)                   & $\lesssim6$ & $(1-50)\hMpc$ & $\lesssim2\times10^{-2}$ & D    & \cite{marchini:13a} \\
    CMB lensing (Planck)                & $\lesssim6$  & $\gtrsim1\hMpc$   & $\lesssim10^{-2}$   & D    & \cite{marchini:13b, hu:13} \\
    Cluster abundance (Chandra)         & $<0.15$      & $(1-10)\hMpc$     & $<1.3\times10^{-4}$ & HS   & \cite{schmidt:09a, ferraro:10} \\
    Cluster abundance (MaxBCG)          & $0.18, \ 0.25$ & $(1-10)\hMpc$   & $<1.9\times10^{-4}$ & D    & \cite{lombriser:10} \\
    Gravitational redshift of galaxies in clusters & $0.1-0.55$ & $(0.5-10)\hMpc$ & \ldots       & HS   & \cite{wojtak:11} \\
    Cluster density profiles (maxBCG)   & $0.23$       & $(0.2-20)\hMpc$   & $<3.6\times10^{-3}$ & HS   & \cite{lombriser:11b} \\
    Coma gas measurements               & $0.02$       & $(0.1-1)\hMpc$    & $<6\times10^{-5}$   & D/HS & \cite{terukina:13} \\
    Strong gravitational lenses (SLACS) & $0.06-0.36$ & $(1-10)~\textrm{kpc}$ & $<2.5\times10^{-6}$ & HS & \cite{smith:09t} \\
    Solar System                        & $0$ &$\lesssim20~\textrm{au}~/~8~\textrm{kpc}$ & $<8\times10^{-7}$ & D/HS & \cite{hu:07a,lombriser:13c} \\
    Supernova monopole radiation        & $\sim0$ & $\sim200R_{\odot}$    & $\lesssim10^{-2}$    & D/HS & \cite{upadhye:13} \\
    Distance indicators in dwarf galaxies & $\lesssim0.002$ & $\lesssim100R_{\odot}$ & $<5\times10^{-7}$ & D/HS & \cite{jain:12} \\
    \hline
    Relativistic effects in galaxy-clustering & $\lesssim1$ & $\gtrsim200\hMpc$ & $\lesssim10^{-1}$ & D & \cite{lombriser:13a} \\
    21 cm intensity mapping + CMB       & $0.7-2.5$    & $\gtrsim50\hMpc$  & $\lesssim10^{-5}$   & D    & \cite{hall:12} \\
    CMB ISW-lensing bispectrum          & $\lesssim5$  & $\gtrsim40\hMpc$  & $\lesssim10^{-2}$   & D/HS & \cite{hu:12, munshi:14} \\
    Matter bispectrum                   & $\sim0$ & $\lesssim30\hMpc$ & $10^{-6}\lesssim\absfR\lesssim10^{-4}$ & HS & \cite{gilmarin:11} \\
    Stacked phase-space distribution     & $0.2-0.4$ & $(1-20)\hMpc$ & $\lesssim(10^{-6}-10^{-5})$ & HS  & \cite{lam:12a} \\
    Galaxy infall kinematics            & $0.25$ & $(0.5-30)\hMpc$ & $\lesssim(10^{-5}-10^{-4})$ & HS   & \cite{zu:13} \\
    Dwarf galaxies                      & $\sim0$      & $\sim1~{\rm kpc}$ & $\lesssim10^{-7}$   & D/HS & \cite{jain:11, vikram:13} \\
    \hline
    \hline
  \end{tabular}
  \end{center}
\end{table*}


Chameleon models have been constrained using a variety of observations from laboratory to cosmological scales~\cite{hu:07a, jain:12, llinares:12, upadhye:12b, brax:13a, erickcek:13a, upadhye:13, erickcek:13b, lombriser:13c, terukina:13}.
Fig.~\ref{fig:constraints} shows the current bounds inferred on the field amplitude and coupling strength on local~\cite{hu:07a, lombriser:13c}, astrophysical~\cite{jain:12}, and cosmological scales~\cite{terukina:13}.
In particular, the Hu-Sawicki and designer $f(R)$ models ($\omega=0$) have been tested using a range of observables and methods.
A selection of current and prospective constraints on $\absfR$ are summarised in Table~\ref{tab:constraints}, focusing mainly on cosmological results.
Note that due to the difference in the signatures of the Hu-Sawicki and designer $f(R)$ models, for rigour, the $95\%$ confidence level constraints listed from the different analyses are assigned to the specific form of $f(R)$ assumed, i.e., the designer (D) and Hu-Sawicki (HS) models.
Hereby, the Hu-Sawicki case refers to the model specifications with $\alpha=0.5$ (or $n=1$).
A few of the results listed in Table~\ref{tab:constraints} shall shortly be reviewed here.

With increasing $\absfR$, the growth of structure becomes enhanced at late times.
This affects the cosmic microwave background (CMB) through modifications of the integrated Sachs-Wolfe (ISW) effect, gravitational lensing, and the Sunyaev-Zel'dovich (SZ) effect.
With growing $\absfR$, the ISW contribution is reduced, initially yielding a decrease of the temperature anisotropy power spectrum at low multipoles.
Eventually, with increasing modification, the ISW contribution to the temperature field becomes negative, at which point, however, the ISW contribution to the temperature anisotropy power spectrum starts to rise again due to the square in the temperature field. 
While the initial reduction of the ISW contribution is slightly preferred by the data, the enhancement with $\absfR$ after the turning point can be used to place constraints on the model at the order of $\absfR\lesssim10^{-1}$~\cite{song:07, lombriser:10}.
When cross correlating the ISW effect with foreground galaxies, the linear contribution of the temperature field yields an anti-correlation for strong modifications, which can be used to tighten constraints on $\absfR$ by about a factor of 5~\cite{giannantonio:09, lombriser:10}.
With the improved measurement of the high angular multipoles, especially by the \emph{Planck} mission, it has also become possible to constrain $\absfR$ with CMB lensing, tightening constraints by almost one order of magnitude over the galaxy-ISW bounds~\cite{marchini:13a, marchini:13b, hu:13}.

Using the galaxy power spectrum measured by \emph{WiggleZ}, Dossett \etal.~\cite{dossett:14} recently derived a constraint of $10^{-4}\lesssim\absfR\lesssim2\times10^{-1}$ and $\absfR\lesssim10^{-5}$ on the designer model, assuming that the modifications at $z\sim0.6$ are accurately described by linear perturbation theory up to scales of $k=0.1\Mpch$ and $k=0.2\Mpch$, respectively.
Measurements of the galaxy clustering in a sample of luminous red galaxies (LRG) from the Sloan Digital Sky Survey (SDSS) combined with the galaxy-galaxy lensing signal and galaxy velocities obtained from redshift-space distortions were used by Reyes \etal.~\cite{reyes:10} to get a measurement of the $E_G$ parameter~\cite{zhang:07} that provides a robust test of gravity and cancels uncertainties in galaxy bias and the initial amplitude of matter fluctuations, however, only yielding weak constraints on $\absfR$~\cite{lombriser:10}.

The enhanced abundance of massive clusters discussed in \tsec{sec:halomassfunctionlinearhalobias} has been used by Schmidt \etal.~\cite{schmidt:09a}, using \emph{Chandra} X-ray data, and Ref.~\cite{lombriser:10}, using SDSS MaxBCG clusters, to constrain the Hu-Sawicki ($\alpha=0.5$) and the designer model, respectively, at the level of $\absfR\lesssim10^{-4}$.
A prescription for mapping these constraints to different values of $\alpha$ in the Hu-Sawicki model was formulated by Ferraro \etal.~\cite{ferraro:10}.
Similarly, the enhanced abundance affects the stacking of cluster density profiles, which furthermore, shows a signature of a matter pile-up in the infall region due to the late-time enhanced gravitational forces in $f(R)$ gravity.
Weak gravitational lensing measurements around maxBCG clusters were used in Ref.~\cite{lombriser:12} to constrain these effects and place an upper bound of $\absfR\lesssim10^{-3}$ on the model.
Comparable constraints to the ones inferred from the increased abundance have been estimated by Yamamoto \etal.~\cite{yamamoto:10} for redshift-space distortions measured in the LRG sample.

Interesting constraints on modified gravity can also be obtained using the difference in dynamically inferred masses to masses inferred from weak lensing discussed in \tsec{sec:dynamicalmassprofile}.
Refs.~\cite{bolton:06, smith:09t} constrained modifications of gravity measuring the dynamical masses of strong lenses via their stellar velocity dispersions and using the radii of Einstein rings to determine the lensing masses in a sample of elliptical galaxies from the Sloan Lens ACS (SLACS) survey.
In specific, Smith~\cite{smith:09t} derived an upper bound of $\absfR<2.5\times10^{-6}$ on the Hu-Sawicki $f(R)$ model ($\alpha=0.5$), assuming that environmental effects that may shield gravitational force modifications can be neglected.
Schmidt~\cite{schmidt:10} modelled the enhancement in the velocity dispersion in $f(R)$ gravity using Eq.~(\ref{eq:dynamicalmass}), where massive halos recover the results of Newtonian gravity, and pointed out that this modification is also reduced for halos which have more massive halos in their proximity.
This mass estimator was used by Wojtak \etal.~\cite{wojtak:11} to constrain modifications of gravity by disentangling the gravitational redshift of the light emitted by galaxies, which propagates through the cluster and shifts the observed galaxy centre, from the kinetic Doppler effect due to galaxy motions, which broadens the observed velocity distribution.
Their measurement is compatible with the $f(R)$ modifications.
Recently, Terukina \etal.~\cite{terukina:13} have combined gas measurements in the Coma cluster of the X-ray temperature and surface brightness as well as from the SZ effect with lensing measurements around the cluster to infer a constraint on $f(R)$ gravity of $\absfR\lesssim6\times10^{-5}$.

Finally, the currently strongest constraints on $f(R)$ gravity in astronomy are inferred from local and astrophysical tests of gravity (see Fig.~\ref{fig:constraints}).
The requirement on the model to satisfy Solar System constraints and the consequent interpolation of the chameleon field from this high-curvature regime to the low curvature on galactic scales and the environment of the Milky Way, assumed to be the cosmological background, yields a constraint of $\absfR<8\times10^{-7}$~\cite{hu:07a, lombriser:13c}.
Note that for the chameleon models given by Eq.~(\ref{eq:potential}), local tests of the equivalence principle such as conducted by the lunar laser ranging experiment, E\"ot-Wash, or the Cassini mission alone only place weak constraints on the modifications of gravity if assuming that the scalar field in the galactic background is in the minimum of its effective potential.
This corresponds to assuming that the scalar field in our Galaxy or the Solar System satisfies the high-curvature solution discussed in \tsec{sec:chameleonmechanism}, in which case the chameleon field is suppressed proportionally to the ratio between the local and background curvature.  
Constraints from the condition of a locally minimised chameleon field can be inferred by requiring that the background scalar field becomes smaller than the galactic gravitational potential within the characteristic scale of the halo, which can be related to constraints on maximal rotation velocities~\cite{hu:07a}, or similarly, by modelling the matter distribution of the Milky Way, solving the resulting scalar field equation, and tracing the chameleon field from the cosmological background to the location of the Solar System, where the high-curvature solution requires it to be shielded~\cite{lombriser:13c}.
The comparison of shielded and potentially unshielded distance indicators from tip of the red giant branch stars and cepheids, respectively, in a sample of unscreened dwarf galaxies gives a constraint of $\absfR<5\times10^{-7}$~\cite{jain:12}.
Another interesting but weaker astrophysical constraint with an upper bound of $\absfR\lesssim10^{-2}$ can furthermore be inferred from the absence of monopole radiation in core-collapse supernovae~\cite{upadhye:13}.


\section{Outlook} \label{sec:outlook}



With the increasing wealth and quality of observational data that will be collected with future surveys, constraints on the gravitational models can be improved and novel methods for testing scalar-tensor gravity and chameleon models will become feasible.
Prospective constraints on $f(R)$ gravity from a variety of observables and on a wide range of scales have, for instance, been analysed in Refs.~\cite{jain:11, gilmarin:11, lam:12a, hu:12, hall:12, lombriser:13a, vikram:13, zu:13, munshi:14}.
The possibility of using relativistic corrections in horizon-scale galaxy clustering, measured in a multi-tracer analysis to cancel cosmic variance, for inferring constraints on dark energy and modified gravity models, has been explored in Ref.~\cite{lombriser:13a}, finding a prospective bound of $\absfR\lesssim10^{-1}$ on the designer $f(R)$ model.
Hall \etal.~\cite{hall:12} have estimated that the combination of 21~{\rm cm} mapping and CMB data will allow to place a constraint on this field amplitude of order $10^{-5}$ and a modification of the same order should also be detectable in the bispectrum of the dark matter density field~\cite{gilmarin:11}.
The cross correlation of the ISW effect with gravitational lensing generates a signature in the bispectrum of the CMB temperature field.
With enhanced growth at late-times, $f(R)$ gravity suppresses the ISW-lensing cross correlation and modifies the temperature bispectrum.
This effect was used by Hu \etal.~\cite{hu:12} and Munshi \etal.~\cite{munshi:14} to forecast a constraint of $\absfR\lesssim10^{-2}$ for \emph{Planck} results.

The enhanced late-time gravitational forces also increase velocities of unshielded objects.
Signatures of this effect have, for instance, been characterised in the velocity power spectrum of dark matter particles~\cite{li:12c}, for redshift-space distortions~\cite{yamamoto:10, simpson:12, jennings:12, okada:12}, in the stacked phase-space distribution of dark matter around galaxy clusters~\cite{lam:12a, lam:13}, for galaxy infall kinematics~\cite{zu:13}, as well as in the spin-up of galactic halos~\cite{lee:12} and proposed as useful test of gravity with expected constraints of the order of $\absfR\lesssim(10^{-6}-10^{-4})$.
Particularly, the differences between dynamically inferred masses and lensing masses of unshielded astronomical objects (see \tsec{sec:dynamicalmassprofile}) and the equivalence in the shielded case due to environmental or self-shielding effects have been pointed out as promising tests of gravity~\cite{hui:09}.
Zhao \etal.~\cite{zhao:11a} quantified the environmental dependence on the relation between the dynamical and lensing masses of dark matter halos, using an indicator for the environment based on distances to neighbouring halos and divided halos from $N$-body simulations into two different samples, which are either isolated or in high-density environments.
They propose that using their estimator and performing a similar division in observed galaxy samples could yield a smoking gun for modified gravity if a correlation between the environment and the difference of dynamical and lensing mass is found.

Dwarf galaxies in low-density environments are objects of particular interest for the search of chameleon modifications of gravity and to infer constraints on the model.
If the chameleon field amplitude is larger than the potential well of the dwarf galaxy and if its environmental density is low enough that it does not shield it, the chameleon force enhances the rotation curves of gas with respect to the self-screened stars, yields a displacement between the two disk, and furthermore, warps the stellar disk and introduces an asymmetry in its rotation curve~\cite{jain:11}.
These signatures can eventually be used to place constraints on the chameleon modification of the order of the potential wells of the dwarf galaxies, i.e., $\absfR\lesssim10^{-7}$.

Besides the gravitational modifications in the dwarf galaxies that are expected to be strongest if the galaxies reside in voids~\cite{hui:09, jain:11}, Martino and Sheth~\cite{martino:09} also pointed out an increase of the abundance of large voids for attractive extra forces, considering a Yukawa-type modification of gravity.
Li \etal.~\cite{li:12} examined this abundance in $N$-body simulations of $f(R)$ gravity, confirming the result of Ref.~\cite{martino:09}, also finding that halos are less screened near voids and that halos in voids are unscreened.
Clampitt \etal.~\cite{clampitt:12} then analysed void statistics in chameleon gravity using excursion set theory to predict the increase in the number density of large voids and furthermore, used this framework to explore the environmental dependence of void properties.
Thereby, small voids in high-density environments were found to be emptier with faster expanding shells, motivating a clustering analysis of small void tracers in redshift-space as a discriminating test of gravity.
In general, deviations in void properties due to the modification of gravity are stronger than in halo properties, which could potentially yield more powerful tests of gravity with future data than from constraining deviations in halo statistics.


It is important to note that in order to exploit the modified gravity effects in these observables for tests of gravity, in addition to running $N$-body simulations, efficient modelling techniques such as based on excursion set theory or semi-analytic approaches as discussed in \tsec{sec:chameleoncosmology} need to be further developed to allow for smooth variations in the model parameter space and to properly account for parameter degeneracies in the signatures.
These modelling frameworks become even more substantial as the parameter space grows with the additional degrees of freedom introduced by the modifications of gravity and emulator approaches become unfeasible.

Ultimately, the techniques described in \tsec{sec:chameleoncosmology} may potentially be generalised to describe the nonlinear large-scale structure formed in the full Horndeski theory described by Eq.~(\ref{eq:horndeski}).
Important advances in this direction have already been realised.
The full linear cosmological perturbations of the Horndeski theory have been derived~\cite{defelice:11} and implemented in a Boltzmann linear theory solver~\cite{hu:13b}.
Note that the quasistatic perturbations take a relatively simple form and can easily be determined by the modifications of gravity in the background.
On nonlinear scales, in parallel to the description of the cosmological small-scale structure formed in the chameleon model, semi-analytic methods have also been developed for other subclasses of the Horndeski action such as for the Galileon and symmetron models~\cite{barreira:13c, barreira:14, brax:13b, taddei:13}.
Barreira \etal.~\cite{barreira:13c} formulated a spherical collapse model for Galileon models, applying their results in Ref.~\cite{barreira:14} to model the halo mass function, halo bias, and halo model matter power spectrum, and finding good agreement with the corresponding statistics extracted from the $N$-body simulations performed in Ref.~\cite{barreira:13b, li:13}.
Similarly, Schmidt \etal.~\cite{schmidt:09c} studied the spherical collapse model in the Dvali-Gabadadze-Porrati (DGP) model~\cite{dvali:00}, which reduces to a Galileon model in the decoupling limit, and used it to determine halo properties, comparing them to $N$-body simulations conducted in Ref.~\cite{schmidt:09b}.
Note that strong constraints on DGP and Galileon models have been inferred using a variety of linear cosmological observations in Refs.~\cite{fang:08a, lombriser:09, raccanelli:12, xu:13} and~\cite{appleby:12, barreira:13a, barreira:14}, respectively, limiting the impact of viable deviations from the concordance model on nonlinear scales.
Finally, the spherical collapse model, halo mass function, halo bias, and nonlinear matter power spectrum has also been analysed for symmetron models~\cite{brax:13b, taddei:13}, a further subclass of the Horndeski action (see \tsec{sec:chameleonmodels}).

While the development of these modelling techniques is still at an early stage, the success in reproducing the important features of the modified gravity models in the nonlinear cosmological structure observed in $N$-body simulations of the models promises an application of these results not just to the Horndeski action.
It may further provide hints for extending generalised linear cosmological perturbation formalisms such as the effective field theory of cosmic acceleration~\cite{defelice:11, bloomfield:12} or the PPF framework~\cite{baker:12}, which both embed the linear perturbations of the Horndeski theory, to nonlinear scales.
Similarly, the success of the description of the nonlinear structure in Galileon and DGP gravity based on these semi-analytic models may also be shared by massive gravity~\cite{derham:14}, providing an interesting area of study for future work.


\section{Conclusion} \label{sec:conclusion}


The presence of a chameleon field in our Universe may modify the gravitational interactions on cluster scales while recovering general relativity locally.
Whereas $N$-body simulations provide an essential tool for studying the effects of the scalar-tensor modification of gravity, especially in regions where the modification is getting suppressed due to the chameleon mechanism, semi-analytic modelling techniques become a necessity for the comparison of the theoretical signatures to observational data.
These tests require an efficient interpolation and extrapolation of the simulated results between different choices of cosmological parameters and model specifications of the gravitational theories of interest.
Emulator approaches will not be feasible for this task as the parameter space grows with the extra degrees of freedom introduced by the different modified gravity models and the computational methods need to be generalised accordingly.

This article summarises a range of observable signatures in the nonlinear cosmological structure that are characteristic for the chameleon modification.
It reviews and compares different techniques to model these observables based on $N$-body simulations, different phenomenological formalisms, fitting functions to simulations, analytical and numerical approximations, the spherical collapse model, excursion set theory, the halo model, and perturbation theory.
Hereby, a particular focus is given to the well studied Hu-Sawicki and designer models of $f(R)$ gravity, for which a summary of the current state of observational constraints is provided and supplemented with an outlook on prospective constraints and novel methods for testing the chameleon modifications.

The semi-analytic methods discussed here are still at an early stage of construction and need to be developed further.
Their success in reproducing the important features of the nonlinear cosmological structure observed in $N$-body simulations of chameleon $f(R)$ gravity, along with similar achievements for describing the cosmological structure in Galileon and symmetron models, anticipates that a generalisation of these modelling techniques and an application thereof to the full Horndeski theory of scalar-tensor gravity may be feasible.
Hence, at the nearing of the 100th anniversary since the formulation of the foundations of general relativity~\cite{einstein:16}, the study of the cosmological structure formed in the simplest extension of the Theory of Gravity, i.e., for scalar-tensor models, and the observational constraints that can be inferred on these extensions promise to remain a very interesting and active field of research.


\paragraph{Acknowledgements}


The author thanks Philippe Brax, Baojiu Li, Patrik Valageas, and Gong-Bo Zhao for sharing numerical results which have been used to produce Figs.~\ref{fig:hmf}, \ref{fig:matterpower1}, and \ref{fig:matterpower2}.
This work has been supported by the STFC Consolidated Grant for Astronomy and Astrophysics of the University of Edinburgh.
Numerical computations have been performed with Wolfram $Mathematica~9$.
Please contact the author for access to research materials.


\bibliographystyle{JHEP}
\bibliography{chamrev}


\end{document}